\documentclass[10pt,journal,compsoc]{IEEEtran}
\ifCLASSOPTIONcompsoc
  \usepackage[nocompress]{cite}
\else
  \usepackage{cite}
\fi

\usepackage{amsmath}
\usepackage{graphicx}
\usepackage{balance}
\usepackage{multirow}
\usepackage{amsmath}
\usepackage{caption}
\usepackage{subcaption}
\usepackage{stfloats}
\usepackage{color}

\ifCLASSINFOpdf
\else
\fi
\begin{document}
%
% paper title
% Titles are generally capitalized except for words such as a, an, and, as,
% at, but, by, for, in, nor, of, on, or, the, to and up, which are usually
% not capitalized unless they are the first or last word of the title.
% Linebreaks \\ can be used within to get better formatting as desired.
% Do not put math or special symbols in the title.
\title{Performance Analysis of\\Network Coding with IEEE 802.11 DCF \\in Multi-Hop Wireless Networks}
%
%
% author names and IEEE memberships
% note positions of commas and nonbreaking spaces ( ~ ) LaTeX will not break
% a structure at a ~ so this keeps an author's name from being broken across
% two lines.
% use \thanks{} to gain access to the first footnote area
% a separate \thanks must be used for each paragraph as LaTeX2e's \thanks
% was not built to handle multiple paragraphs
%
%
%\IEEEcompsocitemizethanks is a special \thanks that produces the bulleted
% lists the Computer Society journals use for "first footnote" author
% affiliations. Use \IEEEcompsocthanksitem which works much like \item
% for each affiliation group. When not in compsoc mode,
% \IEEEcompsocitemizethanks becomes like \thanks and
% \IEEEcompsocthanksitem becomes a line break with idention. This
% facilitates dual compilation, although admittedly the differences in the
% desired content of \author between the different types of papers makes a
% one-size-fits-all approach a daunting prospect. For instance, compsoc 
% journal papers have the author affiliations above the "Manuscript
% received ..."  text while in non-compsoc journals this is reversed. Sigh.

\author{Somayeh~Kafaie,~\IEEEmembership{Student~Member,~IEEE,}
        Mohamed~Hossam~Ahmed,~\IEEEmembership{Senior~Member,~IEEE,}
        Yuanzhu~Chen,~\IEEEmembership{Member,~IEEE,}
        and~Octavia~A.~Dobre,~\IEEEmembership{Senior~Member,~IEEE}% <-this % stops a space
\IEEEcompsocitemizethanks{\IEEEcompsocthanksitem S. Kafaie, M. H. Ahmed, and O. A. Dobre are with Department of Electrical and Computer Engineering, Faculty of Engineering and Applied Science, Memorial University, St. John's, NL A1B 3X5, Canada. \protect\\
Email: \{somayeh.kafaie, mhahmed, odobre\}@mun.ca
\IEEEcompsocthanksitem Y. Chen is with Department of Computer Science, Memorial University, St. John's, NL A1B 3X5, Canada.\protect\\
% note need leading \protect in front of \\ to get a newline within \thanks as
% \\ is fragile and will error, could use \hfil\break instead.
E-mail: yzchen@mun.ca
}% <-this % stops an unwanted space
%\thanks{Manuscript received April 19, 2005; revised August 26, 2015.}
}

\IEEEtitleabstractindextext{%
\begin{abstract}
Network coding is an effective idea to boost the capacity of wireless networks, and a variety of studies have explored its advantages in different scenarios. However, there is not much analytical study on throughput and end-to-end delay of network coding in multi-hop wireless networks considering the specifications of IEEE 802.11 Distributed Coordination Function. In this paper, we utilize queuing theory to propose an analytical framework for bidirectional unicast flows in multi-hop wireless mesh networks. We study the throughput and end-to-end delay of inter-flow network coding under the IEEE 802.11 standard with CSMA/CA random access and exponential back-off time considering clock freezing and virtual carrier sensing, and formulate several parameters such as the probability of successful transmission in terms of bit error rate and collision probability, waiting time of packets at nodes, and retransmission mechanism. Our model uses a multi-class queuing network with stable queues, where coded packets have a non-preemptive higher priority over native packets, and forwarding of native packets is not delayed if no coding opportunities are available. Finally, we use computer simulations to verify the accuracy of our analytical model. 
\end{abstract}
%In the packet forwarding process, a node in our model forward native packets without imposing any extra delay if it is its turn to transmit and there exists no coding opportunity. 

% Note that keywords are not normally used for peerreview papers.
}

% make the title area
\maketitle

% To allow for easy dual compilation without having to reenter the
% abstract/keywords data, the \IEEEtitleabstractindextext text will
% not be used in maketitle, but will appear (i.e., to be "transported")
% here as \IEEEdisplaynontitleabstractindextext when the compsoc 
% or transmag modes are not selected <OR> if conference mode is selected 
% - because all conference papers position the abstract like regular
% papers do.
\IEEEdisplaynontitleabstractindextext
% \IEEEdisplaynontitleabstractindextext has no effect when using
% compsoc or transmag under a non-conference mode.

% For peer review papers, you can put extra information on the cover
% page as needed:
% \ifCLASSOPTIONpeerreview
% \begin{center} \bfseries EDICS Category: 3-BBND \end{center}
% \fi
%
% For peerreview papers, this IEEEtran command inserts a page break and
% creates the second title. It will be ignored for other modes.
\IEEEpeerreviewmaketitle

\IEEEraisesectionheading{\section{Introduction}\label{sec:introduction}}
% Computer Society journal (but not conference!) papers do something unusual
% with the very first section heading (almost always called "Introduction").
% They place it ABOVE the main text! IEEEtran.cls does not automatically do
% this for you, but you can achieve this effect with the provided
% \IEEEraisesectionheading{} command. Note the need to keep any \label that
% is to refer to the section immediately after \section in the above as
% \IEEEraisesectionheading puts \section within a raised box.

% The very first letter is a 2 line initial drop letter followed
% by the rest of the first word in caps (small caps for compsoc).
% 
% form to use if the first word consists of a single letter:
% \IEEEPARstart{A}{demo} file is ....
% 
% form to use if you need the single drop letter followed by
% normal text (unknown if ever used by the IEEE):
% \IEEEPARstart{A}{}demo file is ....
% 
% Some journals put the first two words in caps:
% \IEEEPARstart{T}{his demo} file is ....
% 
% Here we have the typical use of a "T" for an initial drop letter
% and "HIS" in caps to complete the first word.
\IEEEPARstart{C}{apacity} is a crucial resource in multi-hop wireless networks as it is shared not only between the source and destination of data packets but also among relay nodes forwarding packets. To increase the transmission capacity of wirless networks, the powerful concept of network coding~\cite{NC-Ahlswede-IEEETransactionsIT2000} has been introduced, which can improve performance significantly in theory, without considering PHY/MAC layer constraints such as contention, collision and interference. However, network protocols inevitably deal with such physical phenomena and constraints. Therefore, more theoretical studies are needed to better quantify the benefits of network coding over traditional forwarding for actual protocols considering PHY/MAC layer specifications.

There have been many experimental studies on this subject, but much fewer mathematical analyses. Some previous theoretical studies are designed for saturated queues, where each node always has a packet to transmit that would cause an infinite delay. In many cases, researchers consider a simple topology, where source is one~\cite{AnaNC-basic-Sagduyu-2006, AnaNC-basic-Sagduyu-MILCOM2007, oneHop1-Moghadam-CL2016,oneHop2-Moghadam-CL2016} or two~\cite{Ana-NC-2hop-Paschos-inforcom2013, AnaNC-basic-Amerimehr-WC2014, TwoWay2-Jamali-TWC2015, Ana-NC-2hop-Zeng-PDS2014, Ana-NC-2hop-Le-MC2010, PNC2hop-Lin-TWC2013, TwoHop-Lin-TWC2016, TwoWayAna-Amerimehr-ICC2009, Ana2hop-Umehara-SAComm2009, Ana2hop-Umehara-AdhocNet2011} hops away from the destination. Furthermore, the theoretical research on multi-hop networks~\cite{AnaNCQ-Sagduyu-IT2008, AnaNCQ-Iraji-WCNC2009, AnaNCQ-Amerimehr-MC2014, physicalMultiHop-Lin-TWC2016, AnaNCQunicast-Hwang-Net2011, Ana-Kafaie-GlobeCom2016} usually models network coding with simplifying assumptions, such as conflict-free scheduled access, no interference, no collision, or no exponential back-off. Moreover, most research in this subject investigates only the throughput of the network, and postpone the transmission of native packets in favor of providing more coding opportunities.

The main contributions of this paper are as follows: 
\begin{enumerate}
\item We apply multi-class queuing network to study the performance of multi-hop wireless mesh networks applying inter-flow network coding~\cite{COPE-Katti-IEEEACMTransactions2008}, where intermediate nodes can mix packets of different flows by bitwise \emph{XOR} operation. This model provides an analytical framework for a multi-hop chain topology with bidirectional unicast flows in opposite directions. In contrast to other studies, no artificial delay is injected in forwarding native packets even if there is no coding opportunity. In fact, we do not postpone transmission of native packets artificially to generate coded packets (i.e., opportunistic coding). Also, we consider separate classes of queues for native and coded packets, while the coded queue is a higher-priority queue.

\item We develop our analytical framework for both non-coding and coding schemes in multi-hop wireless networks, and formulate not only the throughput but also the end-to-end delay in a stable network.

\item The proposed model takes into account PHY/MAC layer specifications. It applies random medium access CSMA/CA as in IEEE 802.11 Distributed Coordination Function (DCF) with binary exponential back-off considering clock freezing and virtual carrier sensing as explained in Sections~\ref{subsec:dataLayer} and \ref{subsec:nonCodingServiceTime}. We consider retransmission, collision probability, link qualities and coding probabilities in calculating the throughput and an upper-bound of average end-to-end delay of the network. Also, the validity of the analytical model is shown by simulations in NS-2.
\end{enumerate}

The rest of this paper is organized as follows. Related work is discussed in Section \ref{sec:relatedWork}. We explain the system model and assumptions in Section \ref{sec:systemOverview}. Section \ref{sec:problemFormulation} introduces our derived formulation of the throughput and end-to-end delay for the non-coding and coding schemes. To show the accuracy of our analytical model, we compare the results with computer simulations in Section \ref{sec:performanceEvaluation}. Finally, Section \ref{sec:conclusion} draws conclusions, and discusses directions for future work. 

\section{Background and Related Work} \label{sec:relatedWork}

\begin{table*}[!t]
\caption{An overview of the analytical research in the literature.}
\label{table:background}
\centering
\begin{tabular}{|c||c||c||c||c||c||c||c||c||c|}
\hline
Research & Network & Number & \multirow{2}{*}{Throughput} & \multirow{2}{*}{Delay} & Stable & Random & Unicast & Opportunistic & Exponential \\ 
& coding & of hops & & & queues & access & /multidcast & coding & back-off  \\ 
\hline
\cite{AnaNC-basic-Sagduyu-2006, AnaNC-basic-Sagduyu-MILCOM2007} & $\surd$& 1 & $\surd$ & - & $\surd$ &$\surd$ & multicast & - & - \\ 
\hline
\cite{oneHop1-Moghadam-CL2016,oneHop2-Moghadam-CL2016} & $\surd$ & 1 & $\surd$ & - & $\surd$ & $\surd$ & multicast & $\surd$ & - \\ 
\hline
\cite{Ana-NC-2hop-Paschos-inforcom2013} & $\surd$ & 2 & $\surd$ & - & $\surd$ & - & unicast & - & -\\
\hline
\cite{AnaNC-basic-Amerimehr-WC2014} & $\surd$ & 2 & $\surd$ & relay & $\surd$ & - & unicast & both & - \\
\hline  
\cite{TwoWay2-Jamali-TWC2015} & $\surd$ & 2 & $\surd$ & $\surd$ & $\surd$ & $\surd$ & unicast & - & - \\
\hline
\cite{Ana-NC-2hop-Zeng-PDS2014, TwoHop-Lin-TWC2016, TwoWayAna-Amerimehr-ICC2009} & $\surd$ & 2 & $\surd$ & - & $\surd$ & $\surd$ & unicast & - & - \\
\hline
\multirow{2}{*}{\cite{Ana-NC-2hop-Le-MC2010}} & \multirow{2}{*}{$\surd$} & \multirow{2}{*}{2} & \multirow{2}{*}{$\surd$} & \multirow{2}{*}{-} & \multirow{2}{*}{$\surd$} & priority/ & \multirow{2}{*}{unicast} & \multirow{2}{*}{both}&\multirow{2}{*}{-} \\
 &  &  &  &  &  & equal access &  &  & \\
\hline
\cite{PNC2hop-Lin-TWC2013} & $\surd$ & 2 & $\surd$ & - & $\surd$ & $\surd$ & unicast & - & $\surd$ \\
\hline
\cite{Ana2hop-Umehara-SAComm2009, Ana2hop-Umehara-AdhocNet2011} & $\surd$ & 2 & $\surd$ & $\surd$ & $\surd$ & $\surd$ & unicast & $\surd$ & - \\
\hline
\cite{AnaNCQ-Sagduyu-IT2008} & $\surd$ & $\geq 3$ & $\surd$ & - & - & $\surd$ & multicast & - & - \\
\hline
\cite{AnaNCQ-Amerimehr-MC2014} & $\surd$ & $\geq 3$ & $\surd$ & - & $\surd$ & $\surd$ & multicast & - & - \\
\hline
\cite{physicalMultiHop-Lin-TWC2016} & $\surd$ &  $\geq 3$ & $\surd$ & - & - & $\surd$ & unicast & - & - \\
\hline
\cite{AnaNCQunicast-Hwang-Net2011} & $\surd$ & $\geq 3$ & $\surd$ & - & - & $\surd$ & unicast & - & $\surd$  \\
\hline
\cite{DelayAna-Ko-TMC2016} & - & $\geq 3$ & $\surd$ & $\surd$ & $\surd$ & $\surd$ & multicast & not applicable & $\surd$ \\
\hline
 \end{tabular}
\end{table*}

Prior theoretical studies on network coding usually consider a simple topology. Most of them study the performance for a two-way relay~\cite{Ana-NC-2hop-Paschos-inforcom2013, AnaNC-basic-Amerimehr-WC2014, TwoWay2-Jamali-TWC2015, Ana2hop-Umehara-SAComm2009}, or derive some analytical bounds for a single relay in a two-hop region, where multiple sources initiate unicast sessions to multiple destinations~\cite{Ana-NC-2hop-Zeng-PDS2014, Ana-NC-2hop-Le-MC2010, PNC2hop-Lin-TWC2013, TwoHop-Lin-TWC2016, TwoWayAna-Amerimehr-ICC2009, Ana2hop-Umehara-AdhocNet2011}. In particular, Amerimehr and Ashtiani~\cite{AnaNC-basic-Amerimehr-WC2014} study the throughput and delay of a two-way relay by adopting frequency division duplexing (FDD). Without focusing on PHY/MAC layer constraints, they compare the throughput and delay in the relay for two cases where 1) the relay postpones transmission of native packets, and 2) native packets are sent immediately.

Sagduyu et al. study the stable throughput when one or two sources broadcast their packets to two destinations~\cite{AnaNC-basic-Sagduyu-2006} or more~\cite{AnaNC-basic-Sagduyu-MILCOM2007} via independent channels. Paschos et al.~\cite{Ana-NC-2hop-Paschos-inforcom2013} study a two-way relay in inter-flow network coding taking into account overhearing, where coding decisions at the relay are either stochastic or deterministic via receiving overhearing reports. Moghadam and Li~\cite{oneHop1-Moghadam-CL2016,oneHop2-Moghadam-CL2016} study the maximum stable throughput in single-hop wireless networks, where a source multicasts data packets to several destinations directly, and network coding is applied to retransmit the packets not received by a subset of the destinations.

In addition, Jamali et al. propose a dynamic scheduling based on a threshold on the amount of information at nodes' transmission buffers in bidirectional relay networks. This scheduling is used to maximize throughput both without any constraint on the delay~\cite{TwoWay1-Jamali-TWC2015}, and with constraint to guarantee a certain average delay~\cite{TwoWay2-Jamali-TWC2015}. Furthermore, Umehara et al.~\cite{Ana2hop-Umehara-SAComm2009} analyze the throughput and delay of network coding in two-hop networks with two unbalanced traffic cases (i.e., one-to-one and one-to-many bidirectional relay) employing slotted ALOHA. They also extend the model to single-relay multi-user wireless networks and provide the achievable region in throughput~\cite{Ana2hop-Umehara-AdhocNet2011}.

In another single-relay research, Lin et al.~\cite{PNC2hop-Lin-TWC2013} study the throughput of network-layer and physical-layer network coding under IEEE 802.11 DCF with two groups of nodes communicating with each other via a relay node. In a similar work, where again all nodes are in carrier sensing range of each other, they not only study the throughput under slotted ALOHA but also propose a hybrid network coding scheme (i.e., a combination of physical-layer and network-layer network coding) to improve performance~\cite{TwoHop-Lin-TWC2016}.

Regarding multi-hop wireless networks, Sagduyu et al.~\cite{AnaNCQ-Sagduyu-IT2008} consider a collision-free scheduled access to formulate throughput for both saturated and non-saturated queues. However, in case of a random access scheme, their analytical model is limited to saturated queues. In this paper, instead of limiting nodes to scheduled access, we study the performance using IEEE 802.11 MAC layer, where collision can occur without assuming saturated queues. In addition, we provide simulation results to verify our model.

In a similar theoretical-based approach for multicast sessions, Amerimehr et al.~\cite{AnaNCQ-Amerimehr-MC2014} derive throughput for multi-hop wireless networks. They also define a new metric, \emph{network unbalance ratio}, which identifies the amount of unbalance in stability among nodes. However, their estimate of service time does not take into account some important features of IEEE 802.11 DCF like binary exponential random back-off. Furthermore, they postpone transmission of the native packet at a node until receiving a packet from another flow to be combined with it, and thus, causing a long delay. 

In another work considering IEEE 802.11 DCF, Lin and Fu~\cite{physicalMultiHop-Lin-TWC2016} investigate the throughput capacity of physical-layer network coding in which a common center node exchanges packets with others in multi-hop wireless networks. They analyze such canonical networks both with equal and variable link-length, and find the optimal number of hops to maximize the throughput. In addition, Ko and Kim~\cite{DelayAna-Ko-TMC2016} study the throughput and end-to-end delay of multi-hop wireless networks utilizing IEEE 802.11 DCF only for traditional forwarding, when every node initiates a flow with the same rate to a random destination, and same arrival rate is assumed at all nodes. They derive a delay-constrained capacity in terms of carrier sensing range and packet generation rate.

Furthermore, Hwang et al.~\cite{AnaNCQunicast-Hwang-Net2011} propose an analytical framework for bidirectional unicast flows in multi-hop wireless networks. Their work considers collision and different interference levels in CSMA/CA by varying the carrier-sensing range and signal-to-interference ratio to maximize the throughput in different retransmission schemes. In comparison, the novelty of our work consists of the following: 1) our model formulates not only throughput but also end-to-end delay; 2) in our model, opportunistic coding is applied (i.e., if a node has a transmission opportunity, it does not delay forwarding native packets to generate coded packets); 3) our focus is on more realistic case of stable queues.

Table~\ref{table:background} presents an overview of all studies discussed in this section.

\begin{table}[!t]
\caption{The description of used symbols.}
\label{table:symbols}
\centering
\begin{tabular}{|l||l|}
\hline
\textbf{Symbol} & \textbf{Description}\\  \hline
\multirow{2}{*}{$p_{i,j}$} & probability of successful transmission \\
& from $N_{i}$ to $N_{j}$\\ \hline
$\delta$ & maximum propagation delay \\ \hline
$\gamma_{i}$ & packet generation rate at the source $N_{i}$	\\ \hline
$\lambda_{i}$ & arrival rate at $N_{i}$  \\ \hline
$\mu$ & service rate of the queue \\ \hline 
$L_{p}$ & packet length \\ \hline
$\text{CW}_{\min}$ & minimum contention window \\ \hline
\multirow{2}{*}{$\beta$} & maximum number of transmissions \\
& of a packet at each node\\ \hline
$T_\text{data}$ & transmission time of a packet \\ \hline
$T_\text{ack}$ & transmission time of an acknowledgement \\ \hline
$\theta$ & throughput \\ \hline
DIFS & Distributed Inter-Frame Space \\ \hline
SIFS & Short Inter-Frame Space \\ \hline
$T_\text{trans}$ & $T_\text{data}+T_\text{ack}+\text{SIFS}$ \\ \hline
$T_\text{counter}(m)$ & $\text{DIFS}+\dfrac{2^{m-1}\text{CW}_{\min}-1}{2}$ \\ \hline
\multirow{3}{*}{$T(m)$} & time spent on DIFS and back-off in $m^{th}$\\
& transmission by taking into account \\
&the ``clock freezing'' behavior \\ \hline
$W$ & upper-bound of the average end-to-end delay \\ \hline
\multirow{2}{*}{$W^{(j)}$} & upper-bound of the average end-to-end delay\\
& of the $j^{th}$ flow \\ \hline
$T_{s}(m)$ & service time at the $m^{th}$ transmission of a packet \\ \hline
$\lambda_{in,i}^{n(j)}$ & arrival rate of native packets of the $j^{th}$ flow at $N_{i}$\\ \hline
$\lambda_{out,i}^{n(j)}$ & output rate of native packets of the $j^{th}$ flow at $N_{i}$ \\ \hline
$\lambda_{in,i}^{c(j)}$ & arrival rate of coded packets of the $j^{th}$ flow at $N_{i}$\\ \hline
$\lambda_{out,i}^{c}$ & output rate of coded packets at $N_{i}$ \\ \hline
$\lambda_{i}^{n(j)}$ & arrival rate of the $j^{th}$ flow in $Q^n$ of $N_{i}$\\ \hline
$\lambda_{i}^{c}$ &  arrival rate in the coded queue of $N_{i}$	\\ \hline
$W(Q)$ & average waiting time in queue $Q$ \\ \hline
$W_{\text{system}}$ & average waiting time in the queuing system \\ \hline
\multirow{2}{*}{$\bar{R}_i$} & mean residual service time \\
& in priority queues at $N_i$\\ \hline
$\mu_i^{n,\text{seen}}$ & service time seen by lower priority queue, $Q^n$ \\ \hline
\multirow{2}{*}{$P_{\text{mtc}}(r)$} & probability that a packet from flow $r$\\
& in $Q^n$ moves to $Q^c$ \\ \hline
\multirow{2}{*}{$N(r,w)$} & average number of the packets of flow $r$ \\
& arrived in $Q^n$ during $w$ time window \\ \hline
\multirow{2}{*}{$N(r)$} & average number of the packets of flow $r$ \\
& ahead of the currently arrived packet in $Q^n$ \\ \hline
$\pi_{0}(Q^{n(r)})$ & probability of having no packets from flow $r$ in $Q^n$ \\ \hline
\multirow{2}{*}{$I_i$} & set of all nodes in interference range of $N_i$\\
& including $N_i$ \\ \hline
\multirow{2}{*}{$h_x$} &  probability that node $x$ transmits a packet \\ 
& during transmission between two other nodes \\ \hline
\multirow{2}{*}{$P_{i,j}^d$} & probability that $N_i$ drops a packet \\ 
& with next-hop $N_j$ after failure in $\beta$ transmissions \\ \hline
\multirow{2}{*}{$P_{i,j}^\text{decode}$} & decoding probability of a coded packet \\
& arrived at $N_j$ from $N_i$ \\ \hline
 \end{tabular}
\end{table}

\section{System Overview} \label{sec:systemOverview}
Before further explanation of the model, let us summarize the symbols used in this paper in Table~\ref{table:symbols}.
\subsection{Network Model and Assumptions}
\begin{figure}[ht]
\centering
\includegraphics[scale=0.65]{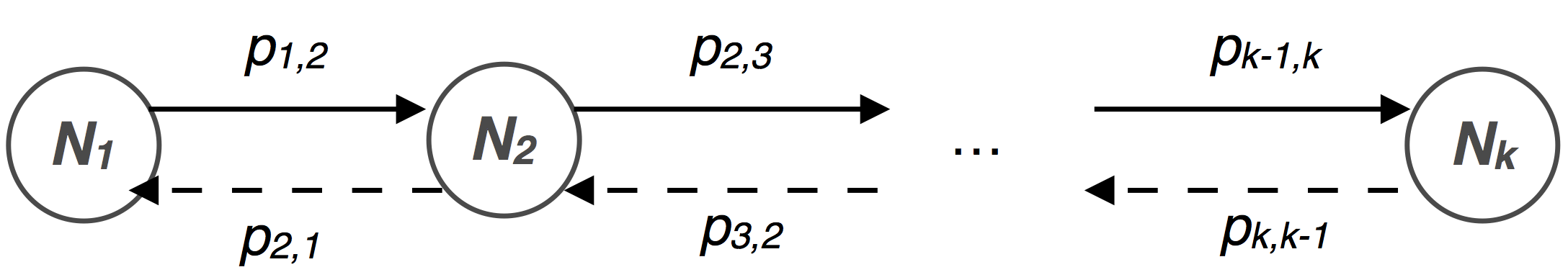}
\caption{Chain topology used for the analytical model.}
\label{fig:chain}
\end{figure}

We propose an analytical model of network coding for bidirectional unicast flows in multi-hop wireless mesh networks to study the throughput and end-to-end delay. Our analytical results are provided for a chain topology with $k$ nodes as depicted in \figurename~\ref{fig:chain}, with two flows in opposite directions. As shown in this figure, $N_{1}$ and $N_{k}$ transmit their packets to each other via intermediate nodes $N_{2}$ to $N_{k-1}$, while we assume that only $N_{i-1}$ and $N_{i+1}$ are in the transmission range of $N_{i}$. 

In this model, we assume that nodes usually do not move, and packets of UDP flows from a source traverse multiple wireless hops to be received by the destination. 
Furthermore, we assume that a node cannot transmit and receive at the same time and the feedback channel is reliable; thus if a node does not hear an acknowledgement (ACK) on time, it assumes that the data packet is lost.

In this network, we consider each node as a queuing system, where the packets in the sending buffer are customers of the queue. We assume that the queues have an unlimited capacity and are in a stable state, i.e., the arrival rate is less than the service rate. 
When a node finds the channel idle, it sends a packet from the head of its queue.  Hence, each node provides services with one server to the packets in its queue, and \emph{Queuing Theory} can be used to model this network.

Our described network has all the properties of \emph{open Jackson networks}~\cite{Jackson-OR57, Jackson-MS63} including 1) each node is considered a queuing system; 2) the packet generation rate 
at source $N_{i}$ ($i=1, k$) follows the Poisson model with a mean rate $\gamma_{i}$;
3) service time at node $N_{i}$ ($i=1,..., k$) is assumed independent from that of other nodes, and it is exponentially distributed with parameter $\mu_{i}$; and 4) a packet that has completed service at node $N_{i}$ (i.e., the packet has been transmitted) will go next to node $N_{j}$ with probability $r_{i,j}$. This probability, presented in (\ref{eq:routingProb}), for the next-hop equals successful transmission probability $p_{i,j}$, and for other nodes equals zero.

\begin{equation}
\label{eq:routingProb}
r_{i,j} =
\left\{
	\begin{array}{ll}
		p_{i,j}  & \mbox{if } N_{j}  \mbox{ is a neighbor of }  N_{i}\\
		0 &  \mbox{elsewhere}
	\end{array}
\right. 
\end{equation}

To formulate this network, we employ concepts from the probability theory, queuing theory and Jackson networks~\cite{QueueBook-Gross-1998, QueueBook-Kleinrock-vol1, QueueBook-Kleinrock-vol2}. Based on the \emph{Burke's} Theorem~\cite{Burke-OR56}, in a stable stationary queuing system, the departure process of an exponential server is Poisson if the arrival rate follows a Poisson process. Furthermore, the \emph{Jackson's Theorem} states that in the Jackson network each node behaves as if its input were Poisson. Therefore, the arrival rate at other nodes, in addition to the sources, can be considered a Poisson process. We assume that $\lambda_{i}$ denotes the arrival rate at $N_{i}$.

\subsection{Data Link Layer Description} \label{subsec:dataLayer}
In this paper, the same data link layer signaling as IEEE 802.11 DCF~\cite{ieee-std-802.11} is applied, with CSMA/CA random access. At the beginning of each time slot, a node, with a packet to transmit, senses the channel. If the node finds the channel idle for a DIFS (Distributed Inter-Frame Space) period of time, it waits for a random back-off interval to minimize the probability of collision with packets transmitted by other nodes, and then transmits the packet. 
We consider a fixed number of time slots for the transmission time of each packet.

The random back-off interval in DCF is discrete with binary exponential growth. To transmit a new packet, random back-off is uniformly chosen from $[0, \text{CW}_{\min}-1]$, where $\text{CW}_{\min}$ is the minimum contention window. When a packet is retransmitted for the $m^{th}$ time ($m>0$), the contention window range will be extended to $[0, 2^{m}\text{CW}_{\min}-1]$, while $2^{m}\text{CW}_{\min}$ is upper-bounded by $\text{CW}_{\max}$. 

Based on the specifications of the IEEE 802.11 standard, the back-off and the DIFS counters are decremented as long as the channel is sensed idle. As soon as it is sensed busy, the node freezes the state of the clock and stops counting down until sensing the idle channel again. Therefore, although the value of DIFS and selected back-off (i.e., the number of ticks in the counter) are specified, the counter may pause due to another transmission which makes the channel busy. This ``clock freezing'' behaviour needs to be taken into account in calculating the back-off time.

The default feedback mechanism in the DCF is automatic repeat request (ARQ), where an ACK is transmitted by the receiver of the data packet, after a period of time called short inter-frame space (SIFS). Since the SIFS is shorter than the DIFS, no other node will sense the channel idle for a DIFS before the end of the ACK transmission. If the sender of a data packet does not receive an ACK before time-out, it will increase the back-off interval and retransmit the packet.

\subsection{The Probability of Successful Transmission}\label{routing}
We calculate the probability of successful transmission in each link in terms of the bit error rate ($p_e$) and collision. In general, a packet transmission, at the link between $N_{i}$ and $N_{j}$, may fail due to packet error rate or collision ($C_{i,j}$), where error rate of a packet with length $L_{p}$ is approximated as $p_e \times L_{p}$.
Thus, the probability of successful transmission of a packet
from $N_{i}$ to $N_{j}$ can be calculated as:
\begin{align} 
p_{i,j} &= (1 - C_{i,j}) (1 - p_e \times L_{p}) \, .
\end{align}

\begin{figure}[ht]
\centering
\includegraphics[scale=0.6]{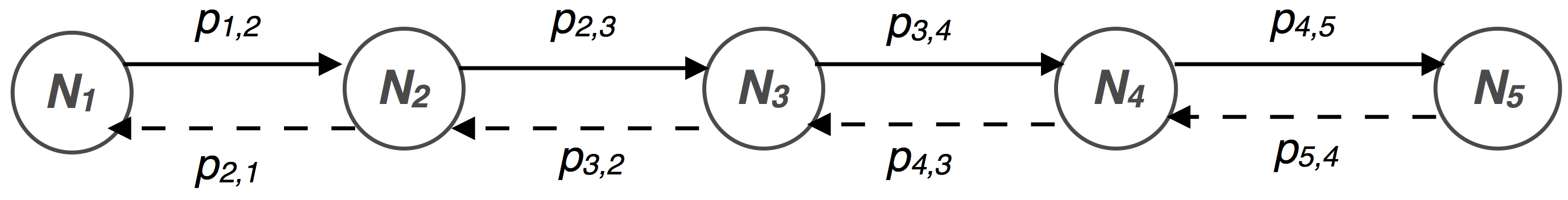}
\caption{Chain topology with 5 nodes.}
\label{fig:chain-5}
\end{figure}

We assume that the probability of collision between a data packet and an ACK is negligible; this is a valid assumption because: 1) the length of ACKs is significantly shorter than the length of data packets, and 2) ACKs are given higher priority and are sent earlier than any data packet. A transmission from $N_i$ to $N_j$ will fail if at the same time, $N_j$ or any other node in its interference range transmits a packet. Let us denote $I_j$ as the set of all nodes in the interference range of $N_j$, including $N_j$ itself. Then the probability of successful transmission from $N_i$ to its neighbor, $N_j$, can be computed as

\begin{equation}
\label{STProb1}
p_{i,j}= (1 - p_e L_{p}) \prod \limits_{N_x \in I_j-\lbrace N_i \rbrace} (1- h_x) \, ,
\end{equation} 
where $h_x$ represents the probability that $N_x$ transmits a packet during packet transmission between two other nodes.

If $N_{i}$ transmits a packet at time $t$, any node in its interference range will sense that the channel is busy after the propagation delay ($\delta$), and avoid any transmission. Therefore, during a propagation delay window before and after $N_{i}$'s transmission (i.e., $(t-\delta, t+\delta)$), other nodes may transmit their packet which will collide with $N_{i}$'s transmission. Although the propagation delay depends on the distance, we assume a fixed propagation delay as the maximum propagation delay. The probability that $N_{x}$ transmits a packet during this time window can be estimated as $h_x = 2 \delta \lambda_{x}$. The proof is given in Appendix A.

As an example, in the chain topology with $5$ nodes depicted in \figurename~\ref{fig:chain-5}, a transmission from $N_{2}$ to $N_{3}$ will fail if at the same time slot that $N_{2}$ is transmitting, $N_{3}$ or $N_{4}$ transmits as well. Note that in this topology, where successive nodes are equally far apart, assuming a two-ray ground reflection propagation model with the default capture threshold of $10$ dB, 
a transmission from $N_{1}$ or $N_{5}$ will not collide with the reception at $N_{3}$ due to capture effect~\cite{TwoRay-Rappaport} (i.e., transmissions from nodes two hops or farther away cannot cause any collision).
Therefore, the probability of a successful transmission from $N_{2}$ to $N_{3}$ equals $p_{2,3} = (1 - 2 \delta \lambda_{3}) (1 - 2 \delta \lambda_{4}) ( 1 - p_e L_p)$. In fact, the following equation can be used to compute the probability of successful transmission from $N_{i}$ to $N_{j}$, when $N_{i}$ and $N_{j}$ are neighbors:
\begin{equation}
\label{STProb2}
p_{i,j}=(1 - p_e L_{p}) \prod \limits_{N_x \in I_j-\lbrace N_i \rbrace} (1-2\delta\lambda_{x}) \, . 
\end{equation}

\section{Problem Formulation} \label{sec:problemFormulation}
In this section, we first provide the analytical model for traditional forwarding (i.e., non-coding scheme); then we extend it to the case that intermediate nodes can utilize network coding and combine packets of the two flows (i.e., coding scheme).
\subsection{Non-coding Scheme}
In the non-coding scheme, the intermediate nodes forward only native packets, while the packets may enter the network (i.e., the queue network) either at node $N_{1}$ with a generation rate $\gamma_{1}$ or at node $N_{k}$ with a generation rate $\gamma_{k}$, and depart from the other end of the chain. Therefore, the intermediate nodes receive packets from both directions. Let $\lambda_{i}^{(1)}$ and $\lambda_{i}^{(2)}$ denote the arrival rate of the first flow (i.e., from $N_{1}$ to $N_{k}$) and the second flow (i.e., from $N_{k}$ to $N_{1}$) arriving at node $N_{i}$, respectively. Therefore, at each node $\lambda_{i}=\lambda_{i}^{(1)} + \lambda_{i}^{(2)}$. 

We consider each node as a single $M/M/1/\infty$ queuing model. As explained earlier, the departure time distribution in an $M/M/1$ queue with arrival rate $\lambda$, in a stable state, is an exponential distribution with mean $1/\lambda$. One of the key rules of probability used in this model states that ``the sum of $t$ independent Poisson processes with arrival rates $\lambda_{1}$, ..., $\lambda_{t}$ is also a Poisson process with an arrival rate $\lambda=\sum\limits_{i=1}^t \lambda_{i}$''~\cite{QueueBook-Gross-1998}. Hence, the assumption of having Poisson arrivals at intermediate nodes holds, and each node can be considered as an independent $M/M/1$ queuing system.

\begin{figure}[ht]
\centering
\includegraphics[scale=0.6]{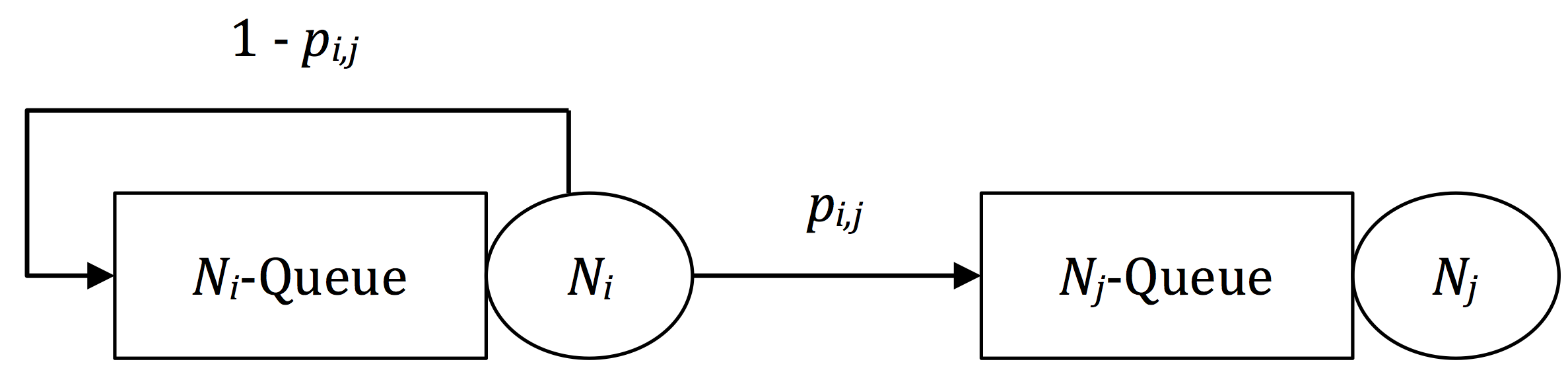}
\caption{Feedback queue to model retransmission.}
\label{fig:feedback}
\end{figure}

To model the retransmission of packets in the network, feedback queues are required. As shown in \figurename~\ref{fig:feedback}, we consider that node $N_{i}$ delivers its packets to the next-hop $N_{j}$ successfully with the probability $p_{i,j}$, and retransmits the packets with the probability $1-p_{i,j}$, at most $\beta - 1$ times (i.e., the packet is retransmitted if the last transmission fails). Hence, a packet is dropped if it cannot be delivered to the next-hop after $\beta$ transmissions. 
This drop probability for a packet sent from $N_i$ to the next-hop $N_j$, can be calculated as
\begin{equation}
\label{eq:Pdrop}
P_{i,j}^d=(1-p_{i,j})^{\beta} \, .
\end{equation}

Taking retransmissions into account, (\ref{eq:lambda4-1}) and (\ref{eq:lambda4-2}) represent the arrival rate of the first flow (i.e., from $N_1$ to $N_k$) and the second flow (i.e., from $N_k$ to $N_1$) at all nodes, respectively.

\begin{subequations}
\label{eq:lambda4}
\begin{equation}
\label{eq:lambda4-1}
\left\{
	\begin{array}{ll}
	   \lambda_{i}^{(1)}=\gamma_{i} + \lambda_{i}^{(1)} (1-p_{i,i+1})(1-P_{i,i+1}^d)  & \mbox{if } i=1\\
		\lambda_{i}^{(1)}= \lambda_{i-1}^{(1)} p_{i-1,i} + \\ ~~~~~~~~~~~\lambda_{i}^{(1)} (1-p_{i, i+1}) (1-P_{i,i+1}^d) & \mbox{if } 1< i < k\\
		\lambda_{i}^{(1)}= \lambda_{i-1}^{(1)} p_{i-1,i}  & \mbox{if } i=k\\
	\end{array}
\right.
\end{equation}
\begin{equation}
\label{eq:lambda4-2}
\left\{
	\begin{array}{ll}
		\lambda_{i}^{(2)}=\gamma_{i} + \lambda_{i}^{(2)} (1-p_{i,i-1})(1-P_{i,i-1}^d)  & \mbox{if } i=k\\
		\lambda_{i}^{(2)}= \lambda_{i+1}^{(2)} p_{i+1,i} + \\ ~~~~~~~~~~~\lambda_{i}^{(2)} (1- p_{i, i-1})(1-P_{i,i-1}^d) & \mbox{if } 1 < i < k\\
		\lambda_{i}^{(2)}= \lambda_{i+1}^{(2)} p_{i+1,i} & \mbox{if } i=1
	\end{array}
\right.
\end{equation}
\end{subequations}

\subsubsection{Successful transmission probabilities}
As explained in Section~\ref{routing}, the probability of transmitting a packet successfully can be calculated in terms of the packet arrival rates and propagation delay by solving the following system of non-linear equations:
\begin{equation}
\label{eq:probabilities2-1}
\begin{cases} 
p_{1,2}=(1-p_e L_p)(1-2 \delta \lambda_{2})(1-2 \delta \lambda_{3}) \\ 
...\\
p_{i-1,i}=(1-p_e L_p)(1-2 \delta \lambda_{i})(1-2 \delta \lambda_{i+1}) \\
...\\
p_{k-2,k-1}=(1-p_e L_p)(1-2 \delta \lambda_{k-1})(1-2 \delta \lambda_{k}^{(2)})\\
p_{k-1,k}=(1-p_e L_p)(1-2 \delta \lambda_{k}^{(2)})\\
p_{k,k-1}=(1-p_e L_p)(1-2 \delta \lambda_{k-1})(1-2 \delta \lambda_{k-2}) \\
...\\
p_{i+1,i}=(1-p_e L_p)(1-2 \delta \lambda_{i})(1-2 \delta \lambda_{i-1})\\
...\\
p_{3,2}=(1-p_e L_p)(1-2 \delta \lambda_{2})(1-2 \delta \lambda_{1}^{(1)})\\
p_{2,1}=(1-p_e L_p)(1-2 \delta \lambda_{1}^{(1)}) \, , \\
\end{cases}
\end{equation}
where all $\lambda_{i}$s are functions of $\gamma_{1}$, $\gamma_{k}$, and successful transmission probabilities as described in~(\ref{eq:lambda4}).

\subsubsection{Service time} \label{subsec:nonCodingServiceTime}
The average service time (i.e., $1/\mu$), which is the time until a packet at the head of the transmission queue of $N_{i}$ is delivered to the next-hop $N_{j}$, can be computed as:
\begin{equation}
\label{eq:service3}
\dfrac{1}{\mu}=\sum \limits_{m=1}^{\beta}p_{i,j}(1-p_{i,j})^{m-1} \sum \limits_{n=1}^m T_{s}(n) \, ,
\end{equation}
where $T_{s}(m)$ denotes the service time at the $m^{th}$ transmission of a packet, which is presented by $T_{s}(m)=T(m) + T_\text{data} + \delta + \text{SIFS} + T_\text{ack} + \delta$, $1 \leq m \leq \beta$. $T_\text{data}$ is the transmission time of a packet (we assume the length of the packets is fixed.), $T_\text{ack}$ denotes the transmission time of an acknowledgement, and $T(m)$ is calculated for the $m^{th}$ transmission of a packet in terms of DIFS and back-off time, considering the ``clock freezing'' feature. 

\begin{figure}[ht]
\centering
\includegraphics[scale=0.7]{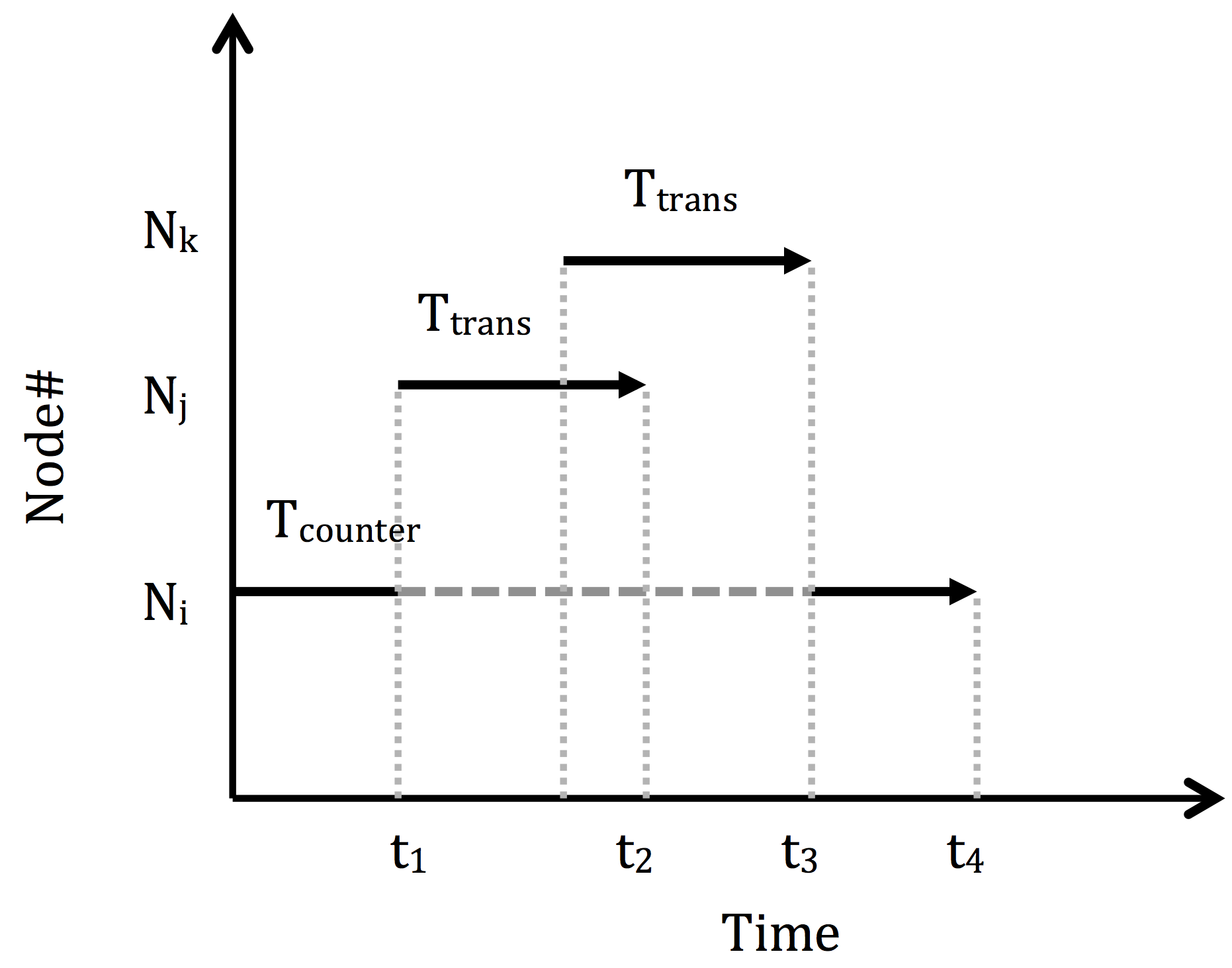}
\caption{Clock freezing behavior of the back-off timer.}
\label{fig:clockFreezing}
\end{figure}

To explain ``clock freezing'', let us use the scenario depicted in \figurename~\ref{fig:clockFreezing}. As shown in this figure, node $N_i$ sets the timer for $T_\text{counter}$ to back-off before transmitting its packet. However at $t_1$, before the back-off timer reaches zero, $N_i$ senses a packet transmission from $N_j$, stops counting down, and freezes the state for $T_\text{trans}=T_\text{data}+T_\text{ack}+\text{SIFS}$. Then, since another transmission by $N_k$ occurs, $N_i$ needs to wait until $t_3$. After that $N_i$ senses the idle channel, resumes the timer, and it is ready to transmit the packet at $t_4$. Note that to consider \emph{network allocation vector (NAV) virtual carrier-sensing} mechanism, we take into account $T_\text{ack}$ and SIFS in calculating $T_\text{trans}$.

To take into account this ``clock freezing'' behavior, we compute the waiting time due to DIFS and back-off as follows
\begin{align} 
\label{eq:back-off1}
T(m) = T_\text{counter}(m) \times e^{-\lambda T_\text{counter}(m)} + \nonumber \\ \sum \limits_{i=1}^{\infty} (T_\text{counter}(m)+i \times T_\text{trans}) \times (1-e^{-\lambda T_\text{trans}})^i \, ,
\end{align}
which means that a node waits $T_\text{counter}$ with the probability that during this period of time, it does not sense any other transmission. In addition, a node waits for $T_\text{counter}+i \times T_\text{trans}$ with the probability that during each $T_\text{trans}$ time period, the node senses at least one packet transmission, and it happens $i$ times. This equation provides an upper-bound for the expected back-off time. Its closed form is calculated as follows, and the proof is given in Appendix B.%\ref{ap:clockFreezing}.

\begin{align} 
\label{eq:back-off2}
T(m) = T_\text{counter}(m) \times e^{-\lambda T_\text{counter}(m)} + \nonumber \\  
(1-e^{-\lambda T_\text{trans}}) \left( \dfrac{T_\text{counter}(m)}{e^{-\lambda T_{\text{trans}}}} + \dfrac{T_\text{trans}}{e^{-2\lambda T_\text{trans}}} \right) \, .
\end{align}

In~(\ref{eq:back-off1}) and (\ref{eq:back-off2}), $\lambda$ represents the sum of arrival rates of the group of nodes which are in carrier sensing range of this node, and the term in the second line of both equations presents the probability of $i$ transmissions from the nodes of this group during the waiting time $T_\text{counter}(m)$. In addition, $T_\text{counter}(m)$ is calculated in terms of $m$, the number of transmissions of a packet, considering the \emph{binary exponential random back-off} interval, as:
\begin{equation}
\label{eq:t-counter}
T_\text{counter}(m)=  \text{ DIFS} + \dfrac{2^{m-1}\text{CW}_{\min}-1}{2},  1 \leq m \leq \beta \, .
\end{equation}

Note that since the random back-off has a uniform distribution, its mean for the $m^{th}$ transmission equals $\dfrac{2^{m-1}\text{CW}_{\min}-1}{2}$.

\subsubsection{Throughput}
It is clear that the throughput, denoted by $\theta$, is identical to the arrival rate at the destinations. Thus, it can be calculated by adding the arrival rate of the second flow at $N_{1}$ and the arrival rate of the first flow at $N_{k}$ as follows
\begin{equation}
\label{eq:throughput2}
\theta= \lambda_{1}^{(2)} + \lambda_{k}^{(1)} \, .
\end{equation}

\subsubsection{End-to-end delay}
The average end-to-end delay equals the summation of the time that each packet spends at the source and intermediate nodes. Also, the time spent at each node consists of the waiting time in the queue, and the time which takes a packet at the head of the queue to be delivered to the next-hop (i.e., service time). Based on queuing theory, the average time a packet spends at node $N_{i}$ until it is delivered to the next-hop, defined as $W_{i}$, can be expressed as   
\begin{equation}
\label{eq:waiting}
W_{i}=\dfrac{1}{\mu_{i}-\lambda_{i}} \, .
\end{equation}

Since in (\ref{eq:service3}) we calculated an upper-bound of the service time (i.e., an upper-bound of $T(m)$), $W_i$ presents an upper-bound of the waiting time at node $N_i$. There are two flows in the network; hence, we calculate the end-to-end delay for the packets of each flow separately, and then we compute the average end-to-end delay by applying the weighted average over the end-to-end delay of the two flows. It is clear that the end-to-end delay for each flow equals the sum of waiting time of the packets of the flow in different nodes, except for the destination. Therefore, an upper-bound of the end-to-end delay for the first and second flows can be computed by~(\ref{eq:WFlow2-1}) and~(\ref{eq:WFlow2-2}), respectively.

\begin{subequations}
\label{eq:WFlow2}
\begin{equation}
\label{eq:WFlow2-1}
W^{(1)}= W_{1}^{(1)} + \sum \limits_{i=2}^{k-1} W_{i}
\end{equation}

\begin{equation}
\label{eq:WFlow2-2}
W^{(2)}=W_{k}^{(2)}+ \sum \limits_{i=2}^{k-1} W_{i} \, ,
\end{equation}
\end{subequations}
where  $W_{i} = \dfrac{1}{\mu_{i}-\left(\lambda_{i}^{(1)}+\lambda_{i}^{(2)}\right)}$ for intermediate nodes. 

Note that while at intermediate nodes the packets of both flows arrive in the queue, in the sources' queue the only packets arrived are those of the flow initiated from that node. Due to this reason, the waiting times at the sources are $W_{1}^{(1)}=\dfrac{1}{\mu_{1}-\lambda_{1}^{(1)}}$ and $W_{k}^{(2)}=\dfrac{1}{\mu_{k}-\lambda_{k}^{(2)}}$. 
Then, an upper-bound of the average end-to-end delay can be computed as
\begin{equation}
\label{eq:eToeDelay2}
W=\dfrac{\gamma_{1}}{\gamma_{1}+\gamma_{k}} \times W^{(1)}+\dfrac{\gamma_{k}}{\gamma_{1}+\gamma_{k}} \times W^{(2)} \, .
\end{equation}

\subsection{Coding Scheme}
To model network coding, we use multi-class queuing networks, and consider that native and coded packets enter separate queues. Furthermore, coded packets in $Q^c$ have a non-preemptive higher priority over the native packets in $Q^n$. This means that a coded packet will be forwarded earlier than all the packets waiting in $Q^n$, but a native packet in service (i.e., the native packet which is being transmitted) is not interrupted by coded packets.

As in the previous case, we assume that the rate of generating packets at $N_{1}$ and $N_{k}$ equals $\gamma_{1}$ and $\gamma_{k}$, respectively, and $\lambda_{i}^{(1)}$ and $\lambda_{i}^{(2)}$ represent the rate of the first and the second flow at $N_{i}$, respectively. Also, we define $\lambda_{i}^{n}$ as the arrival rate of native packets, and $\lambda_{i}^{c}$ as the arrival rate of coded packets at $N_{i}$. 

\subsubsection{Coding module} \label{subsubsec:coding module}
\begin{figure}[ht]
\centering
\includegraphics[scale=0.39]{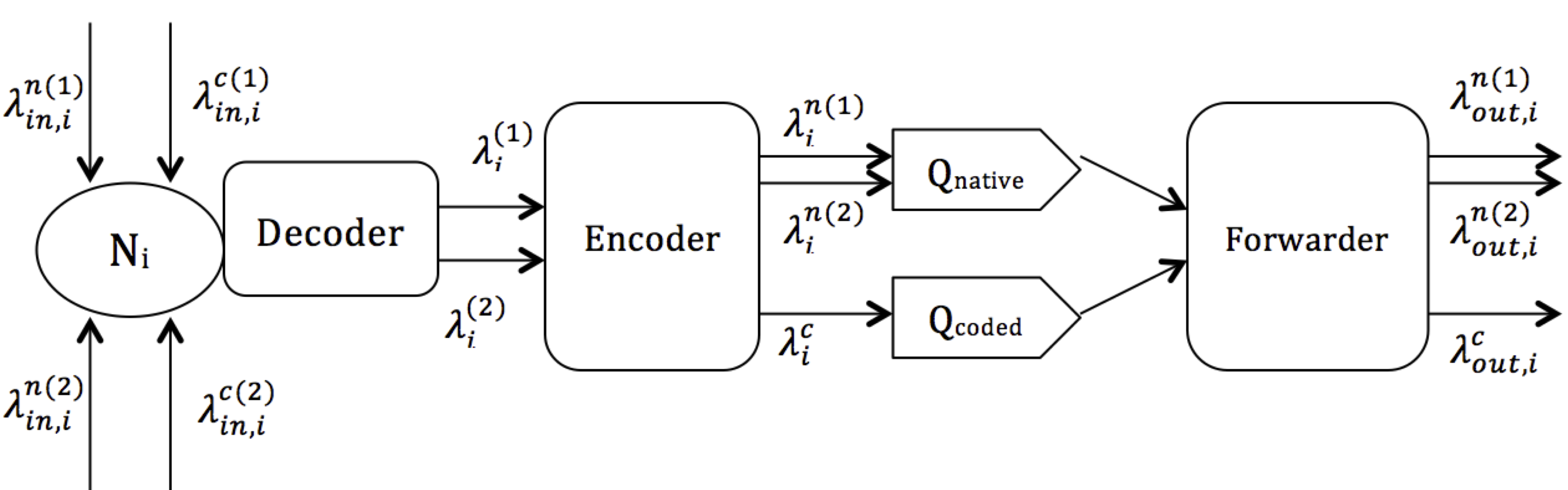}
\caption{A packet from arrival until departure.}
\label{fig:coding-module}
\end{figure}

As shown in \figurename~\ref{fig:coding-module}, $N_{i}$ receives native and coded packets of both flows from the previous hops. Although a coded packet is the combination of both flows, the receiver $N_{i}$ is the next-hop of either the first flow or the second flow (i.e., intended flow). Due to this reason, we distinguish coded packets of different flows arriving at $N_{i}$.

The \emph{decoder}, in \figurename~\ref{fig:coding-module}, decodes the received coded packets and finds the next-hop of the packets. The outputs of this module are native packets of the first and the second flows with rates $\lambda_{i}^{(1)}$ and $\lambda_{i}^{(2)}$, respectively. In fact $\lambda_{i}^{(1)}$ ($\lambda_{i}^{(2)}$) represents the sum of the arrived native packets of the first (second) flow, denoted by $\lambda_{in,i}^{n(1)}$ ($\lambda_{in,i}^{n(2)}$), the successfully decoded packets of the first (second) flow, and retransmitted packets. Therefore, the arrival rates at the \emph{encoder} for both flows (i.e., $\lambda_{i}^{(1)}$ and $\lambda_{i}^{(2)}$ in \figurename~\ref{fig:coding-module}) are calculated as

\begin{subequations}
\begin{equation}
\left\{
	\begin{array}{ll}
	   \lambda_{i}^{(1)} = \lambda_{in,i}^{n(1)} + \lambda_{in,i}^{c(1)} P_{ i-1,i}^\text{decode}+\\ ~~~~~~~~\lambda_{i}^{(1)} (1- p_{i, i+1})(1-P_{i,i+1}^d)    ~~~~~ \mbox{if } 1 \leq i < k\\
	   \lambda_{i}^{(1)} = \lambda_{in,i}^{n(1)} + \lambda_{in,i}^{c(1)}    ~~~~~~~~~~~~~~~~~~~~~~ \mbox{if } i= k\\
	\end{array}
\right.
\end{equation}
\begin{equation}
\left\{
	\begin{array}{ll}
		\lambda_{i}^{(2)} = \lambda_{in,i}^{n(2)} + \lambda_{in,i}^{c(2)} P_{ i+1,i}^\text{decode} +\\ ~~~~~~~~\lambda_{i}^{(2)}  (1-p_{i, i-1})(1-P_{i,i-1}^d) ~~~~~\mbox{if } 1 < i \leq k \\
		\lambda_{i}^{(2)} = \lambda_{in,i}^{n(2)} + \lambda_{in,i}^{c(2)}      ~~~~~~~~~~~~~~~~~~~~~~\mbox{if } i=1 
	\end{array}
\right.
\end{equation}
\end{subequations}

Note that in a general topology, to decode a coded packet with two coding partners, the node should have already received one of them from the opposite direction. Therefore, the decoding probability of a coded packet arrived at $N_i$ from $N_{i-1}$ (or $N_{i+1}$) is $P_{ i-1,i}^\text{decode}=(1-P_{ i+1,i}^d)$ (or $P_{ i+1,i}^\text{decode}=(1-P_{ i-1,i}^d)$).
However, in the chain topology discussed in this paper, the decoding probability is always one (i.e., $P_{ i-1,i}^\text{decode}=P_{ i+1,i}^\text{decode}=1$). In fact, if $N_{i}$ receives coded packet $P_{1} \oplus P_{2}$ from $N_{i-1}$, and $P_{1}$ is its intended packet (i.e., the packet that this node was its next-hop), it must have already received $P_2$ from $N_{i+1}$; otherwise, $N_{i-1}$ could not have received $P_2$ to combine it with $P_1$. 

Previous analytical studies on network coding usually do not consider opportunistic coding, and assume that the transmission of a native packet at the head of the queue, ready to be forwarded, is postponed until receiving packets from other flows, to mix them with the native packet, and send coded packets instead of native ones as much as possible. This assumption provides more coding opportunities, and simplifies estimating the rate of coding opportunities (i.e., forwarding coded packets) at each node. For example, in the chain topology explained here, the rate would be calculated as the minimum of the arrival rates of the flows.
 
However, this postponing will increase the end-to-end delay extremely, especially when the flows are asymmetric as the transmission of native packets should be delayed, waiting for coding partners to arrive. In addition, many practical and well-known network coding protocols are designed based on opportunistic coding, and do not impose such an artificial delay~\cite{COPE-Katti-IEEEACMTransactions2008, DCAR-Le-TMC2010, BEND-Zhang-CNJournal2010}. To limit the delay in the network, and also to analyze the behavior of network coding in practical scenarios, we do not hold transmission of native packets. This means that the arrival rate in $Q^c$ is not the minimum of the arrival rates of the two flows any more, and can be calculated as will be explained here.

In our model, a packet may be transmitted natively if it is at the head of $Q^n$, and there is no packet in $Q^c$. Therefore, the encoder receives the arrived native packet $P$ from flow $r$ ($r={1,2}$), and looks for a packet from flow $\bar{r}$ (i.e., the flow from the opposite direction that can be mixed with flow $r$, $\bar{r}=3-r$) in $Q^n$. If the node finds such a packet $\bar{P}$, it removes $\bar{P}$ from $Q^n$, mixes it with $P$ and adds the coded packet to $Q^c$; otherwise, it will add $P$ to $Q^n$. Therefore, a packet will be added to $Q^c$ if a native packet from flow $r$ arrives at the encoder, and if $Q^n$ contains at least one packet from the flow $\bar{r}$.

On the other hand, packet $P$, from flow $r$, will be sent natively if before it is forwarded, it cannot be mixed with any packet from the other flow. This happens if 1) when it arrives, the queue of the other flow is empty, and 2) during the time that $P$ is waiting in $Q^n$ to be forwarded, the number of packets of the other flow which arrive in $Q^n$ is less than the number of packets of flow $r$ in $Q^n$ ahead of $P$. Note that although all native packets arrive at the same queue, we send them to two separate virtual queues, one for each flow, to be able to calculate the number of packets of each flow in the queue.

If we denote $W(Q^n)$ as the waiting time of an arrived native packet in $Q^n$, then the number of packets of flow $r$ that arrive in $Q^n$ during this time equals $N(r, W(Q^n))=\lambda_{i}^{(r)}\times W(Q^n)$. When $P$ from flow $r$ arrives, if the number of packets of its flow in $Q^n$ (i.e., packets of flow $r$ ahead of $P$) is less than $N(\bar{r}, W(Q^n))$, $P$  is moved from $Q^n$ to $Q^c$ before it is forwarded; otherwise, it stays in $Q^n$. Thus, the probability that a packet from flow $r$ moves to the coded queue of node $N_i$, $Q_i^c$, even if it first arrives in the native queue, $Q_i^n$, can be calculated as

\begin{align} 
\label{eq:moveToCoded1}
P_{\text{mtc}}(r)&= \text{Pr}[N(\bar{r}, W(Q^n)) > N(r)] \nonumber \\ 
&=\sum \limits_{k=0}^{\infty} \text{Pr}[(N(\bar{r}, W(Q^n)) > N(r)) | (N(r)=k)] \nonumber \\& 
\times \text{Pr}[N(r)=k] \nonumber \\
&= \sum \limits_{k=0}^{\infty} \text{Pr}[(N(\bar{r}, W(Q^n)) > k)] \text{Pr}[N(r)=k] \nonumber \\
&=\sum \limits_{k=0}^{\infty} \left( 1 - \sum \limits_{j=0}^k \dfrac{e^{\left(-\lambda_i^{n(\bar{r})}W(Q_i^n\right)} \left( \lambda_i^{n(\bar{r})} W(Q_i^n) \right)^j }{j!} \right)  \nonumber \\&
\left( \dfrac{\lambda_i^{n(r)}}{\mu_i^{n,\text{seen}}}  \right)^k \left( 1 -  \dfrac{\lambda_i^{n(r)}}{\mu_i^{n,\text{seen}}} \right) \, ,
\end{align}
where 
%$\bar{r}$ is the flow from the opposite direction that can be combined with flow $r$, 
$N(r, w)$ denotes the number of packets of flow $r$ arrived in $Q^n$ during time window $w$ and $N(r)$ is the number of the packets of flow $r$ ahead of the currently arrived packet in $Q^n$. Also, $\mu_i^{n,\text{seen}}$ denotes the service time seen by $Q_i^n$ as is discussed later. The closed form of~(\ref{eq:moveToCoded1}) can be computed as
\begin{equation}
\label{eq:moveToCoded2}
P_{\text{mtc}}(r)= 1 - e^{\left( W(Q_i^n)\lambda_i^{n(\bar{r})} \left( \dfrac{\lambda_i^{n(r)}}{\mu_i^{n,\text{seen}}} - 1 \right) \right)} \, .
\end{equation} 
We provide the proof in Appendix C.%~\ref{ap:moveToCoded}.

Next the arrival rate of the packets of the first and second flows in the native queue of $N_i$ is calculated as
\begin{subequations}
\begin{equation}
\lambda_{i}^{n(1)} = \lambda_{i}^{(1)} \times \pi_{0}(Q_i^{n(2)}) \times (1-P_{\text{mtc}}(1)) \, ,
\end{equation} 
\begin{equation}
\lambda_{i}^{n(2)} = \lambda_{i}^{(2)} \times \pi_{0}(Q^{n(1)}) \times (1-P_{\text{mtc}}(2)) \, ,
\end{equation}
\end{subequations}
where $\pi_{0}(Q)$ is the probability that queue $Q$ is empty. This equation means that the arrival rate of native packets of the first flow in $Q_i^n$ (i.e., $\lambda_i^{n(1)}$) equals the arrival rate of the packets of the first flow at the encoder (i.e., $\lambda_i^{(1)}$) for which, in their arrival time, 1) there is no packet from the second flow in $Q_i^n$ (i.e., $\pi_{0}(Q_i^{n(2)})$), and 2) the packet will stay in $Q_i^n$ during its waiting time in the queue (i.e., $1-P_{\text{mtc}}(1)$). $\lambda_{i}^{n(2)}$ is calculated in a similar way. Also, the arrival rate in the coded queue of $N_i$, can be calculated as
\begin{equation}
\lambda_{i}^{c} = \dfrac{\lambda_i^{(1)}+\lambda_i^{(2)}-\lambda_i^{n(1)}-\lambda_i^{n(2)}}{2} \, .
\end{equation} 
The division by two is because each coded packet is a combination of two native packets.

\subsubsection{Native and coded queues}
The arrival rates in $Q_i^n$ and $Q_i^c$ equal $\lambda_{i}^{n(1)}+\lambda_{i}^{n(2)}$ and $\lambda_{i}^{c}$, respectively. The \emph{forwarder} module, in \figurename~\ref{fig:coding-module}, is responsible for forwarding packets. If $Q_i^c$ is not empty, it will select the packet from the head of $Q_i^c$; otherwise, the packet is chosen from the head of $Q_i^n$ if it is not empty.

As stated earlier, priority queues are used to model this case, where the arrival rate in $Q_i^n$ is the sum of the arrival rates of both flows (i.e., $\lambda_{i}^{n}=\lambda_{i}^{n(1)}+\lambda_{i}^{n(2)}$), and the total arrival rate in the queuing system of $N_{i}$ is presented by $\lambda_{i}=\lambda_{i}^{n}+\lambda_{i}^{c}$. Knowing the input rate of native and coded packets at all nodes, one can calculate the output rate at different nodes. Note that since we assume that the queuing system is in a stable state, the departure rates equal the arrival rates ($\lambda_{out,i}^{n(1)}=\lambda_{i}^{n(1)}, \lambda_{out,i}^{n(2)}=\lambda_{i}^{n(2)}, \lambda_{out,i}^{c}=\lambda_{i}^{c}$). Finally, the throughput can be computed using (\ref{eq:throughput2}).

\begin{table}[!t]
\caption{Input rates of native packets at all nodes.}
\label{table:inputRateNative}
\centering
\begin{tabular}{|c||c||c|}
\hline
\boldmath{$i$}&\boldmath{$\lambda_{in,i}^{n(1)}$} & \boldmath{$\lambda_{in,i}^{n(2)}$}\\  \hline
$i=1$ & $\gamma_{1}$ & $\lambda_{out,i+1}^{n(2)} \times p_{i+1,i}$ \\ \hline
$ 1 < i < k$ & $\lambda_{out,i-1}^{n(1)} \times p_{i-1,i}$ & $\lambda_{out,i+1}^{n(2)} \times p_{i+1,i}$\\ \hline
$i=k$ & $\lambda_{out,i-1}^{n(1)} \times p_{i-1,i}$ & $\gamma_{k}$\\ \hline
 \end{tabular}
\end{table}

\begin{table}[!t]
\caption{Input rates of coded packets at all nodes.}
\label{table:inputRateCoded}
\centering
\begin{tabular}{|c||c||c|}
\hline
\boldmath{$i$}& \boldmath{$\lambda_{in,i}^{c(1)}$} & \boldmath{$\lambda_{in,i}^{c(2)}$}\\  \hline
$i=1, i=2$ & $0$ & $\lambda_{out,i+1}^{c} \times p_{i+1,i}$\\ \hline
$ 2 < i < k-1$ & $\lambda_{out,i-1}^{c} \times p_{i-1,i}$ & $\lambda_{out,i+1}^{c} \times p_{i+1,i}$\\ \hline
$i=k, i=k-1$ & $\lambda_{out,i-1}^{c} \times p_{i-1,i}$ & $0$\\ \hline
 \end{tabular}
\end{table}

Tables~\ref{table:inputRateNative} and \ref{table:inputRateCoded} provide the input rates of native and coded packets at all nodes. Moreover, it is clear that the output rate of the first flow at $N_{1}$ and that of the second flow at $N_{k}$ are $\gamma_{1}$ and $\gamma_{k}$, respectively. In addition, the output rate of the second flow and coded packets at $N_{1}$ and the output rate of the first flow and coded packets at $N_{k}$ are equal to zero, as in~(\ref{eq:lambda-out}).
\begin{equation}
\label{eq:lambda-out}
\left\{
	\begin{array}{ll}
	   \lambda_{out,1}^{n(1)}=\gamma_{1} \\
		\lambda_{out,k}^{n(2)}=\gamma_{k} \\
		\lambda_{out,1}^{n(2)}= 0 \\
		\lambda_{out,k}^{n(1)}=0 \\
		\lambda_{out,i}^{c}= 0& \mbox{if } i=1,k
	\end{array}
\right.
\end{equation}

\subsubsection{Service time and end-to-end delay}
As stated earlier, we use two different types of queues for native and coded packets, while the coded packets in $Q^c$ have a non-preemptive higher priority over native packets in $Q^n$. In such a scenario, the service time seen by the native packets is different from the service time of a regular $M/M/1$ queue. The reason is that a native packet at the head of $Q^n$ should wait for all packets in $Q^c$ to be transmitted before its turn for transmission. To estimate the service time \emph{seen} by the native packets (i.e., the packets in lower priority queue) at $N_i$, denoted by $\mu_i^{n, \text{seen}}$, we start from the formula in queuing theory, which calculates the waiting time of a packet in a $M/M/1$ queuing system as

\begin{equation}
W_{\text{system}}=\dfrac{1}{\mu-\lambda} \, .
\end{equation}

Therefore, the service time can be calculated as $\mu=\lambda+1/W_{\text{system}}$. Since for the native queue at $N_i$, $\lambda=\lambda_i^{n(1)}+\lambda_i^{n(2)}$, the waiting time of the packets in the lower priority queue (i.e., $Q_i^n$) can be computed as $W(Q_i^n)=\bar{R}_i/(1-\rho_i^c)(1-\rho_i^c-\rho_i^n)$, and the waiting time of native packets before delivery to the next-hop equals $W_\text{system}=W(Q_i^n)+\dfrac{1}{\mu_i^n}$, we can calculate the service time \emph{seen} by the packets in $Q_i^n$ as 

\begin{align}
\mu_i^{n,\text{seen}}=& \lambda_i^{n(1)}+\lambda_i^{n(2)} + \nonumber \\&
\dfrac{1}{\dfrac{\bar{R}_i}{(1-\rho_i^c)(1-\rho_i^c-\rho_i^n)}+\dfrac{1}{\mu_i^n}} \, ,
\end{align}
where $\mu_i^n$ is the service time of native packets in a regular queuing system that has been calculated earlier in (\ref{eq:service3}). As presented in (\ref{eq:moveToCoded1}) and (\ref{eq:moveToCoded2}), $\mu_i^{n,\text{seen}}$ is used to calculate $P_{mtc}$. Table~\ref{table:formula} shows the required equations to compute variables described in this subsection.

\begin{table}[!t]
\caption{The calculation of some variables' values.}
\label{table:formula}
\centering
\begin{tabular}{|c||c|}
\hline
\textbf{Variable} & \textbf{Equation}\\  \hline
$\rho_{i}$ & $\dfrac{\lambda_{i}}{\mu_{i}}$ \\ \hline 
$\bar{R}_i$ & $\dfrac{\rho_i^n}{\mu_i^n}+\dfrac{\rho_i^c}{\mu_i^c}$ \\ \hline
$W(Q_i^c)$ & $\dfrac{\bar{R}_i}{(1-\rho_i^c)}$ \\ \hline
$W(Q_i^n)$ & $\dfrac{\bar{R}_i}{(1-\rho_i^c)(1-\rho_i^c-\rho_i^n)}$ \\ \hline
$\mu_i^{n,\text{seen}}$ & $\lambda_i^{n(1)}+\lambda_i^{n(2)} + \dfrac{1}{W(Q_i^n)+\dfrac{1}{\mu_i^n}}$ \\ \hline
 \end{tabular}
\end{table}

Furthermore, when a node sends a coded packet, it needs to wait for more than one ACK. In our model with two flows, the service time for coded packets, $\mu_i^c$, is calculated using (\ref{eq:service3}) again, where 
\begin{equation}
\label{eq:Wcoded}
T_{s}(m)=T(m) + T_\text{data} + \delta + 2\times(\text{SIFS} + T_\text{ack} + \delta) \, .
\end{equation} 

Since packets in $Q^c$ and $Q^n$ have different average waiting times, we calculate the waiting time of native and coded packets separately at each node, and then apply the weighted average to compute the average waiting time at each node, as 
\begin{align}
\label{eq:WeToeDelay5}
W_{i} &=\dfrac{\lambda_{i}^{c}}{\lambda_{i}^{c}+\lambda_{i}^{n}} \times \left(W(Q_i^c)+\dfrac{1}{\mu_i^c}\right) \nonumber \\ &+\dfrac{\lambda_{i}^{n}}{\lambda_{i}^{c}+\lambda_{i}^{n}} \times \left(W(Q_i^n)+\dfrac{1}{\mu_i^n}\right) \, .
\end{align}

Finally, the average end-to-end delay can be computed using~(\ref{eq:WFlow2})-(\ref{eq:eToeDelay2}). Note that we assume that the encoding and decoding delays are negligible, and the coding overhead is small enough that we can consider similar length for coded and native packets. In addition, since we calculate an upper-bound of $T_s(m)$, our analytical model provides an upper-bound of the end-to-end delay for both non-coding and coding schemes. 

\section{Performance Evaluation} \label{sec:performanceEvaluation}

\subsection{Network Description}
To verify the accuracy of our proposed analytical model, we run simulations in NS-2 for the chain topology depicted in \figurename~\ref{fig:chain}, where the distance between successive nodes is $200$ m, and $N_{1}$ and $N_{k}$ transmit packets to each other via intermediate nodes $N_{2}$, ..., $N_{k-1}$. The channel propagation used in NS-2 is a two-ray ground reflection model~\cite{TwoRay-Rappaport}, the transmission range is $250$ m, and the carrier sensing range is $550$ m. Hence, in our chain topology, the nodes within two-hop distance of each node are in its carrier sensing range. However, due to the capture effect, the interference range is limited to the nodes one hop away.

In our simulation, we use the IEEE 802.11 standard as the MAC layer protocol, and our physical layer introduces random packet loss by adopting bit error rates ($p_e$). Therefore, the receiver will drop the packet with a probability which is calculated in terms of $p_e$. In addition, a node may drop a packet due to collision. Based on the specifications, a node transmits a packet at most $7$ times (i.e., $\beta=7$).  

The link rate is set to 2 Mbps. The sources, in our simulation scenarios, send Poisson data flows with a datagram size of $1000$ bytes.  We compare the analytical results with the simulation results in different scenarios in terms of throughput and end-to-end delay by varying the packet generation rate and bit error rate. Also, to compare coding and non-coding schemes, we calculate the maximum stable throughput for both cases. 

\subsection{The Effect of Packet Generation Rate} \label{subsec:generationRate}

\begin{figure*}
\centering
\begin{subfigure}{.5\textwidth}
  \centering
  \includegraphics[width=0.95\linewidth]{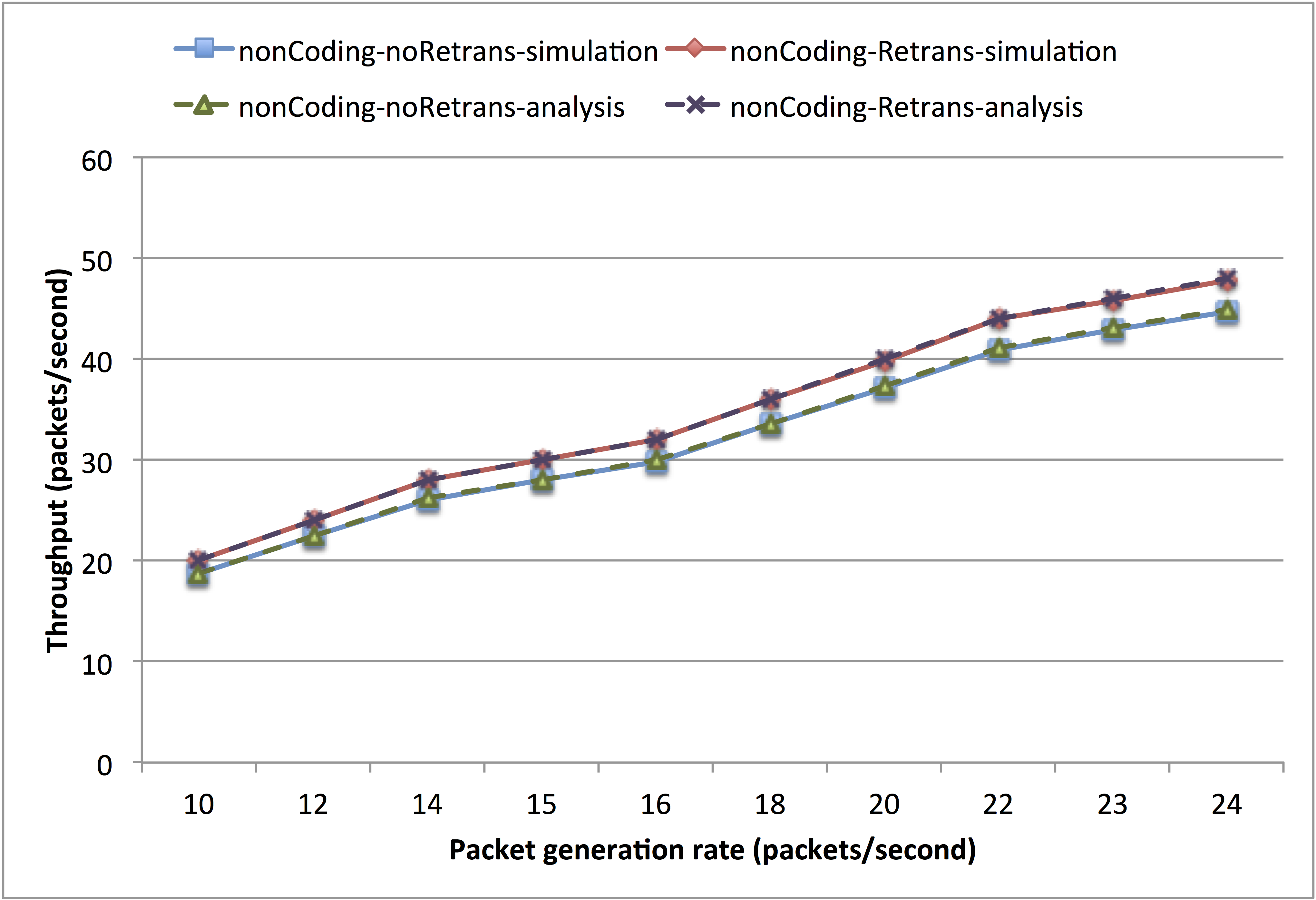}
  \caption{Non-coding scheme.}
  \label{fig:throughput-arrivalRate-nonCoding}
\end{subfigure}%
\begin{subfigure}{.5\textwidth}
  \centering
  \includegraphics[width=0.95\linewidth]{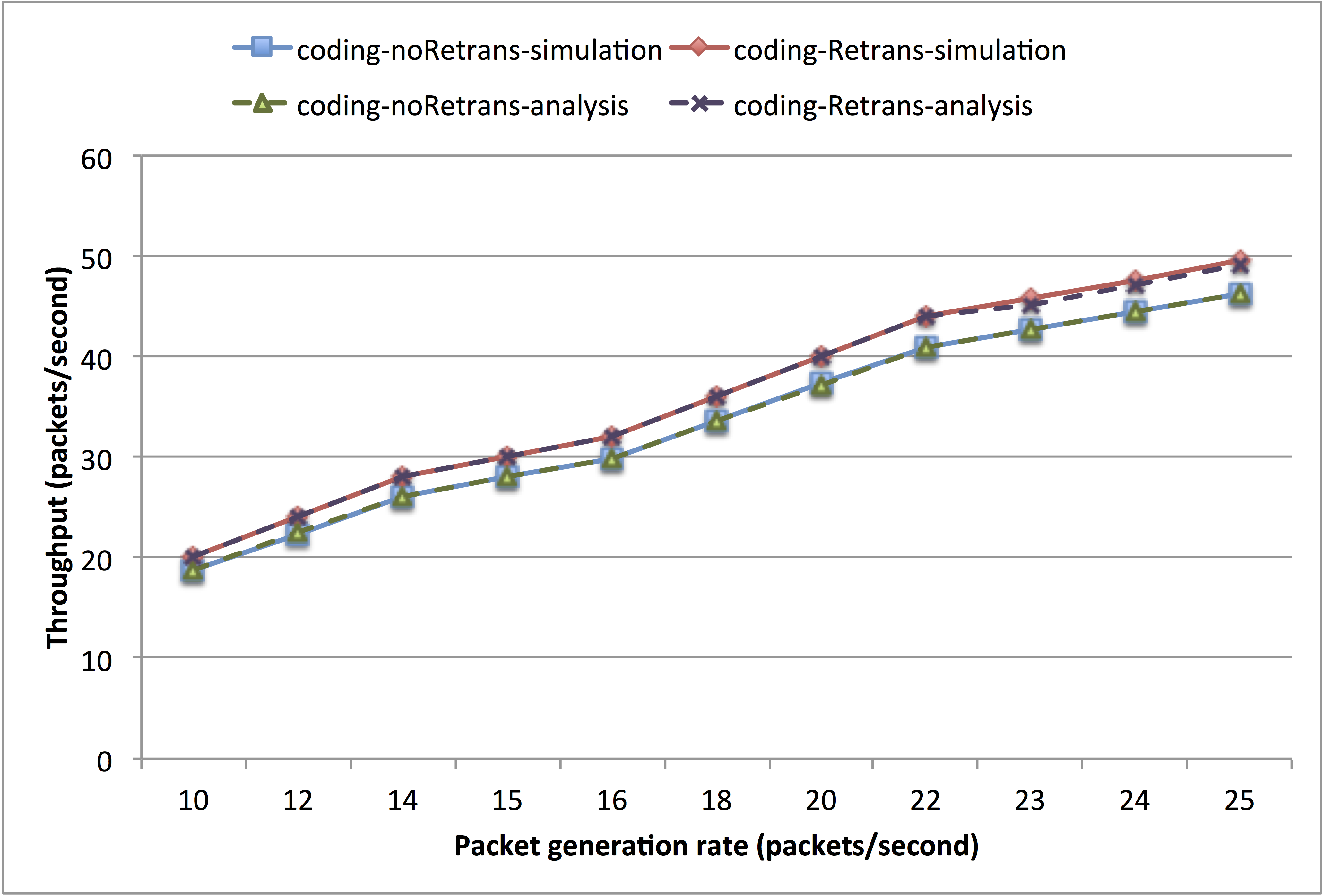}
  \caption{Coding scheme.}
  \label{fig:throughput-arrivalRate-coding}
\end{subfigure}
\caption{Throughput comparison for different packet generation rates in a chain topology with 5 nodes and  $p_e= 2\times 10^{-6}$.}
\label{fig:throughput-arrivalRate}
\end{figure*}

\begin{figure*}
\centering
\begin{subfigure}{.5\textwidth}
  \centering
  \includegraphics[width=0.95\linewidth]{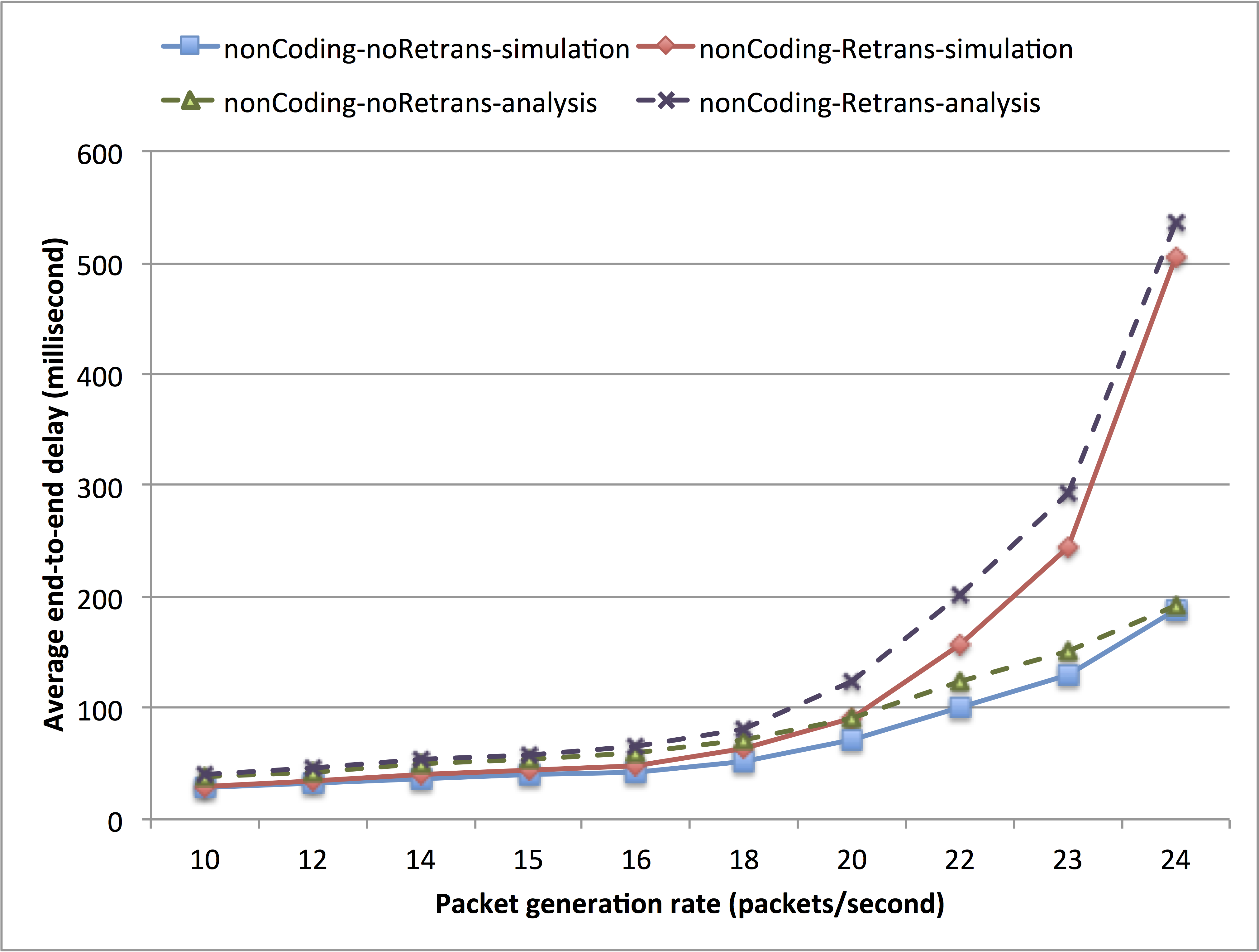}
  \caption{Non-coding scheme.}
  \label{fig:delay-arrivalRate-nonCoding}
\end{subfigure}%
\begin{subfigure}{.5\textwidth}
  \centering
  \includegraphics[width=0.95\linewidth]{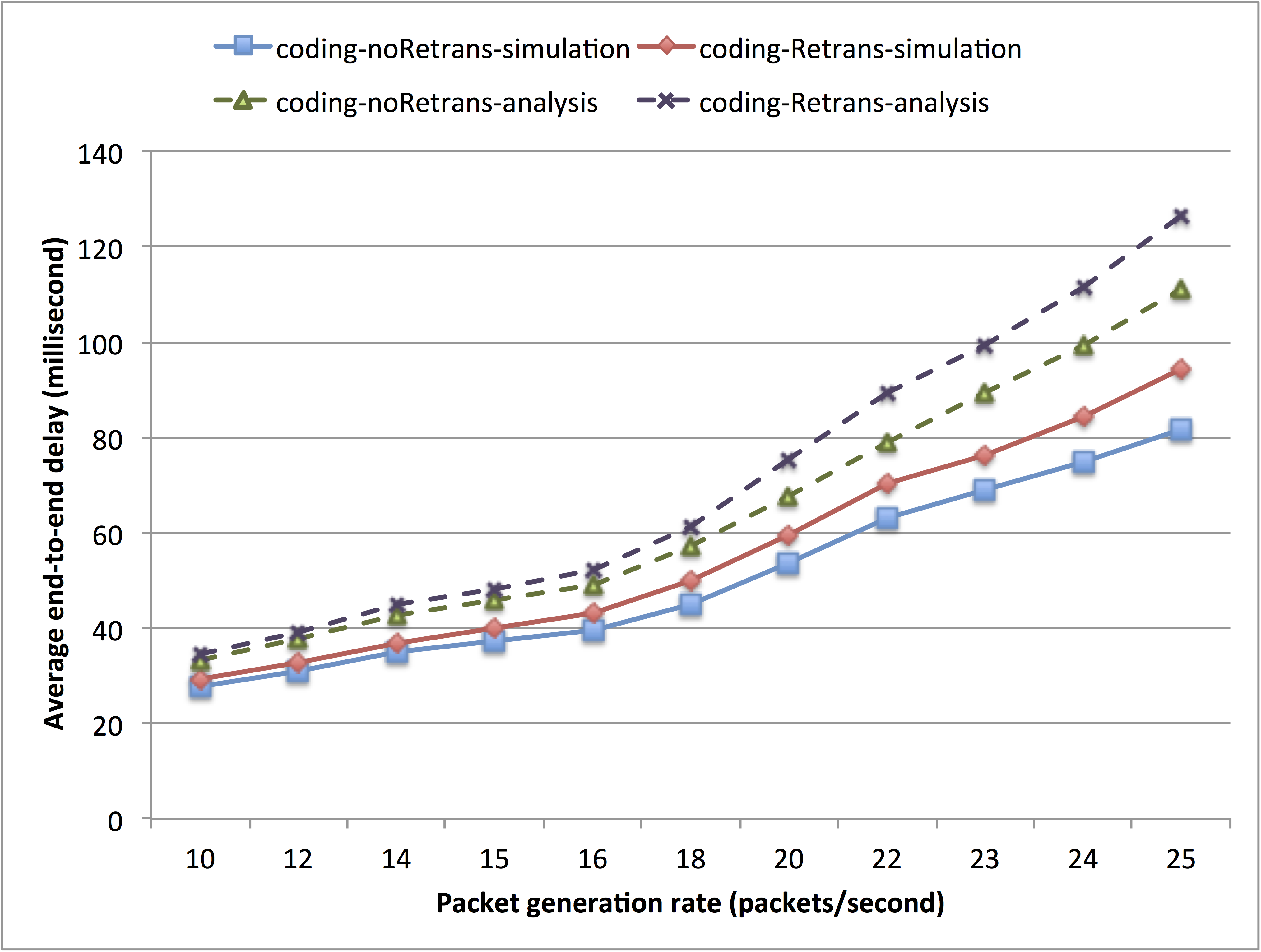}
  \caption{Coding scheme.}
  \label{fig:delay-arrivalRate-coding}
\end{subfigure}
\caption{The average end-to-end delay comparison for different packet generation rates in a chain topology with 5 nodes and $p_e = 2\times 10^{-6}$.}
\label{fig:delay-arrivalRate}
\end{figure*}

In this section, we compare the simulation and analytical results for several packet generation rates in the topology depicted in \figurename~\ref{fig:chain-5} with 5 nodes. In our simulations, Poisson flows between $N_{1}$ and $N_{5}$ last for 170 seconds. We change the generation rate of packets at sources while the bit error rate is fixed to $2 \times 10^{-6}$, and calculate the total throughput and an upper-bound of the average end-to-end delay by assuming an equal packet generation rate at sources (i.e., $\gamma = \gamma_{1}=\gamma_{k}$). We compare the simulation and analytical results for the cases that 1) nodes do not retransmit a packet even if its transmission fails, and 2) nodes transmit a packet at most $\beta$ times ($\beta=7$). 

\figurename~\ref{fig:throughput-arrivalRate}-a presents the analytical and simulation results of throughput for non-coding scheme, both with and without retransmission. Also, \figurename~\ref{fig:throughput-arrivalRate}-b shows the same result for coding scheme. The consistency of the simulation and analytical results corroborates the validity of our analytical model. In addition, one may notice that the throughput at each given packet generation rate is higher when retransmission is enabled. In fact, by disabling the reransmission mechanism, all the efforts to deliver a packet are wasted even if it has made all the way but the very last hop.

Comparing \figurename~\ref{fig:throughput-arrivalRate}-a and \figurename~\ref{fig:throughput-arrivalRate}-b, no considerable throughput gain can be seen for coding scheme in comparison with non-coding scheme especially in lower arrival rates. This is due to the fact that without holding native packets, network coding usually shows its gain over the traditional forwarding approach, where arrival rates are high enough to provide frequent coding opportunities. We will discuss the gain further in Section~\ref{subsec:mst}. 

Regarding the average end-to-end delay, the results in \figurename~\ref{fig:delay-arrivalRate} show that our analytical model provides an upper-bound for the average end-to-end delay in different packet generation rates for both non-coding and coding schemes. In addition, in both scenarios (i.e., with and without retransmission), the average end-to-end delay increases with the packet generation rate; the reason is that at higher generation rates more packets are queued at nodes, which increases the waiting time and consequently the end-to-end delay of the network. However, the end-to-end delay is shorter when retransmission is disabled because each packet has only one transmission chance to be delivered to the next-hop, and lost packets do not contribute to delay calculation. 

As a matter of fact, without retransmission a packet is either dropped or delivered to the next hop with only one transmission. On the other hand, with enabling retransmission, the packet is provided with up to $\beta$ chances to repeat, which improves throughput at the cost of a longer delay. Furthermore, as shown in \figurename~\ref{fig:delay-arrivalRate}, without utilizing network coding the delay grows faster. This is due to the fact that network coding allows more than one packet to be delivered to the next-hop in one transmission, which accelerates packet delivery, and reduces contention.

\subsubsection{Throughput-delay trade-off}
As presented in \figurename~\ref{fig:throughput-arrivalRate}, if the end-to-end delay of the network is finite (i.e., the queues are in stable state), the throughput is an increasing function of packet generation rate~\cite{DelayAna-Ko-TMC2016}.
On the other hand, \figurename~\ref{fig:delay-arrivalRate} shows that the end-to-end delay is also an increasing function of the packet generation rate. As explained earlier, this is due to the fact that generating new packets faster increases the number of packets queued to be transmitted, which means longer waiting time in the queues, as confirmed by (\ref{eq:waiting}).

This verifies a trade-off between throughput and end-to-end delay that has been discussed in the literature~\cite{tradeoffTD-Gamal-TIT2006, DelayAna-Ko-TMC2016, TwoWay2-Jamali-TWC2015}. To find the delay-constraint capacity, one needs to calculate the optimal packet generation rate for a given end-to-end delay. In our model, for both traditional forwarding and network coding, it can be calculated by increasing the packet generation rate as long as the end-to-end delay is less than the given value. Doing so, one can obtain the packet generation rate in which the network achieves the maximum throughput satisfying
the end-to-end delay requirement.

\subsection{The Effect of Bit Error Rate} \label{subsec:BER}

\begin{figure*}
\centering
\begin{subfigure}{.5\textwidth}
  \centering
  \includegraphics[width=0.95\linewidth]{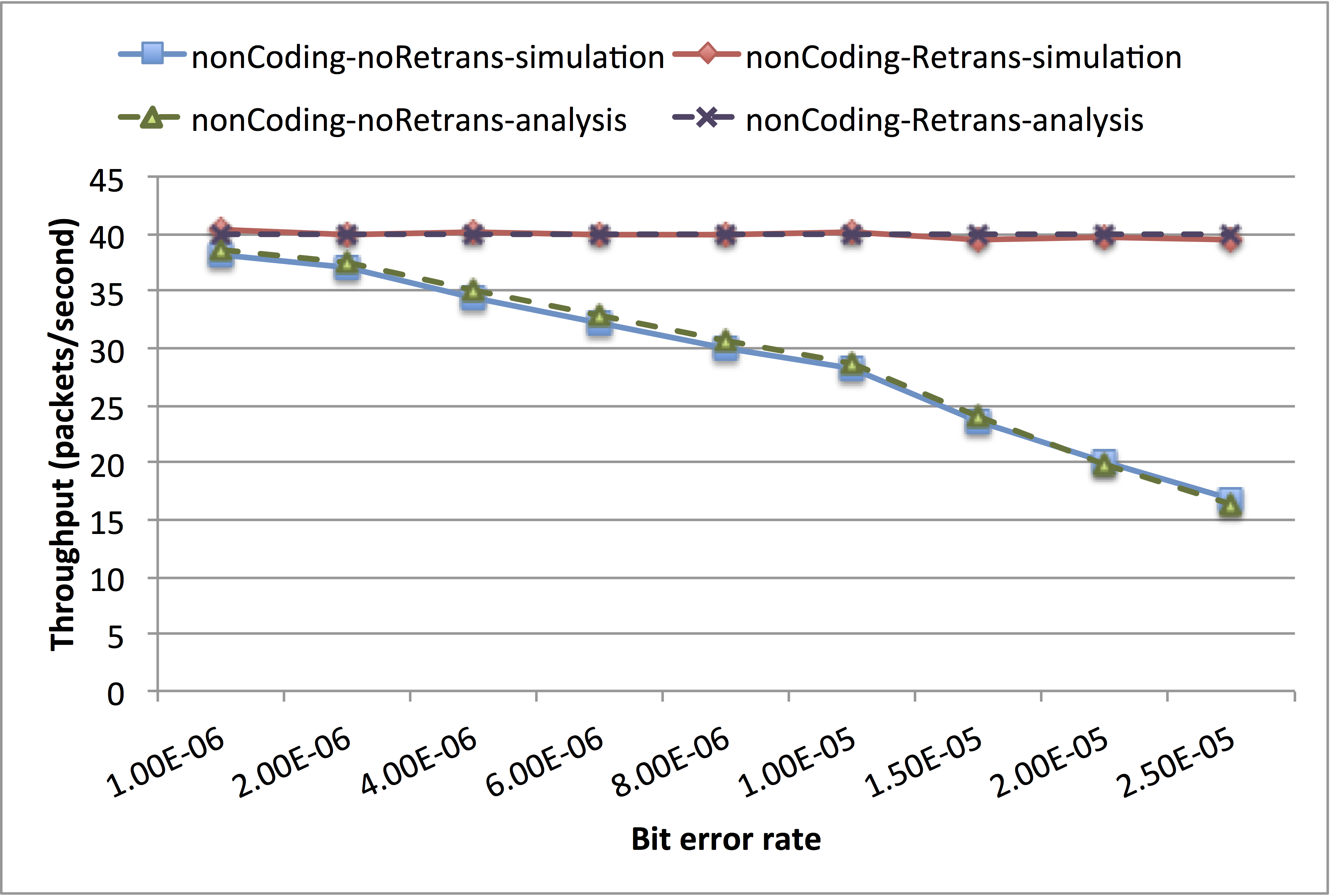}
  \caption{Non-coding scheme.}
  \label{fig:throughput-BER-nonCoding}
\end{subfigure}%
\begin{subfigure}{.5\textwidth}
  \centering
  \includegraphics[width=0.95\linewidth]{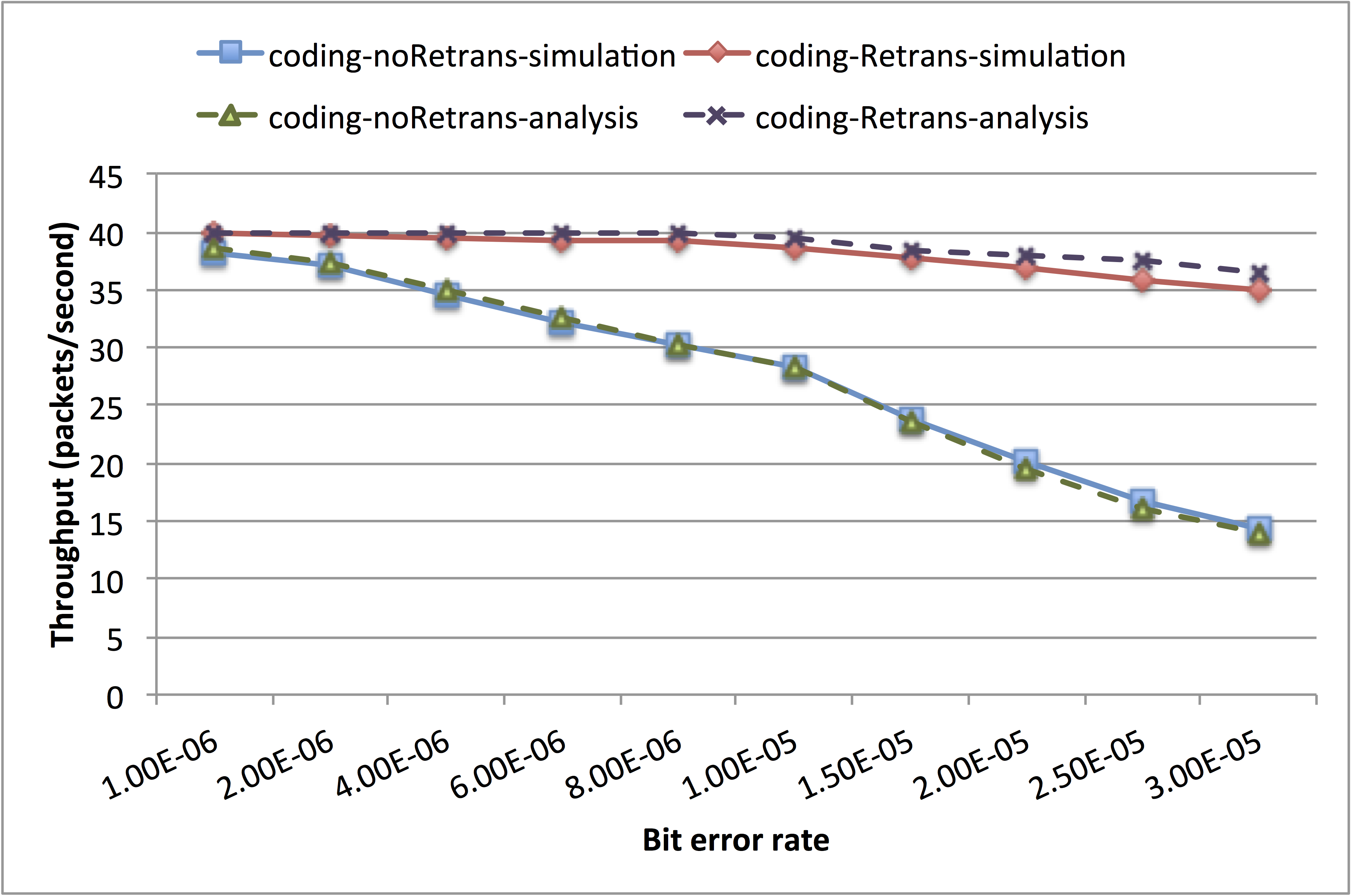}
  \caption{Coding scheme.}
  \label{fig:throughput-BER-coding}
\end{subfigure}
\caption{Throughput comparison for different bit error rates in a chain topology with 5 nodes and $\gamma = 20$.}
\label{fig:throughput-BER}
\end{figure*}
 
\begin{figure*}
\centering
\begin{subfigure}{.5\textwidth}
  \centering
  \includegraphics[width=0.95\linewidth]{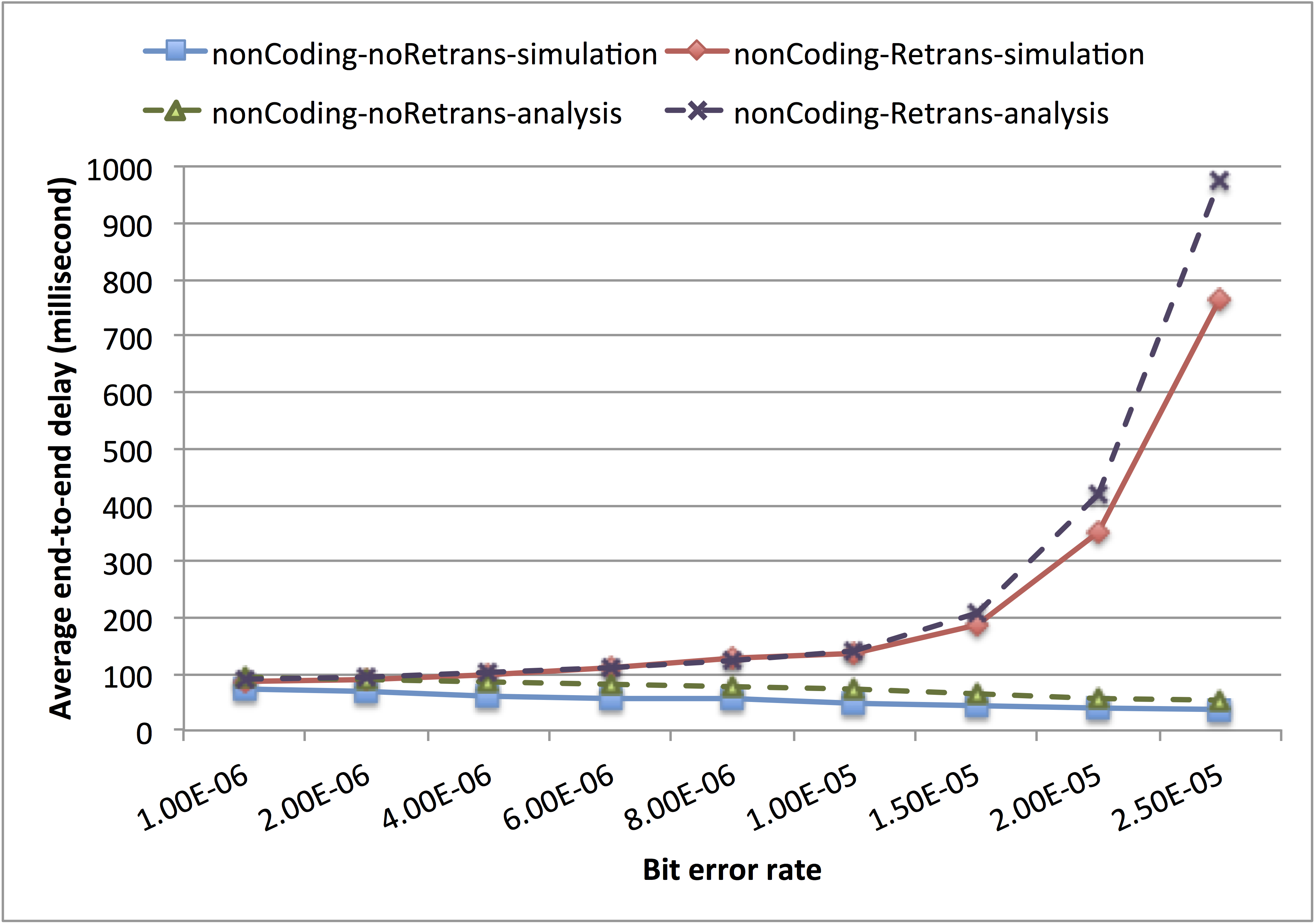}
  \caption{Non-coding scheme.}
  \label{fig:delay-BER-nonCoding}
\end{subfigure}%
\begin{subfigure}{.5\textwidth}
  \centering
  \includegraphics[width=0.95\linewidth]{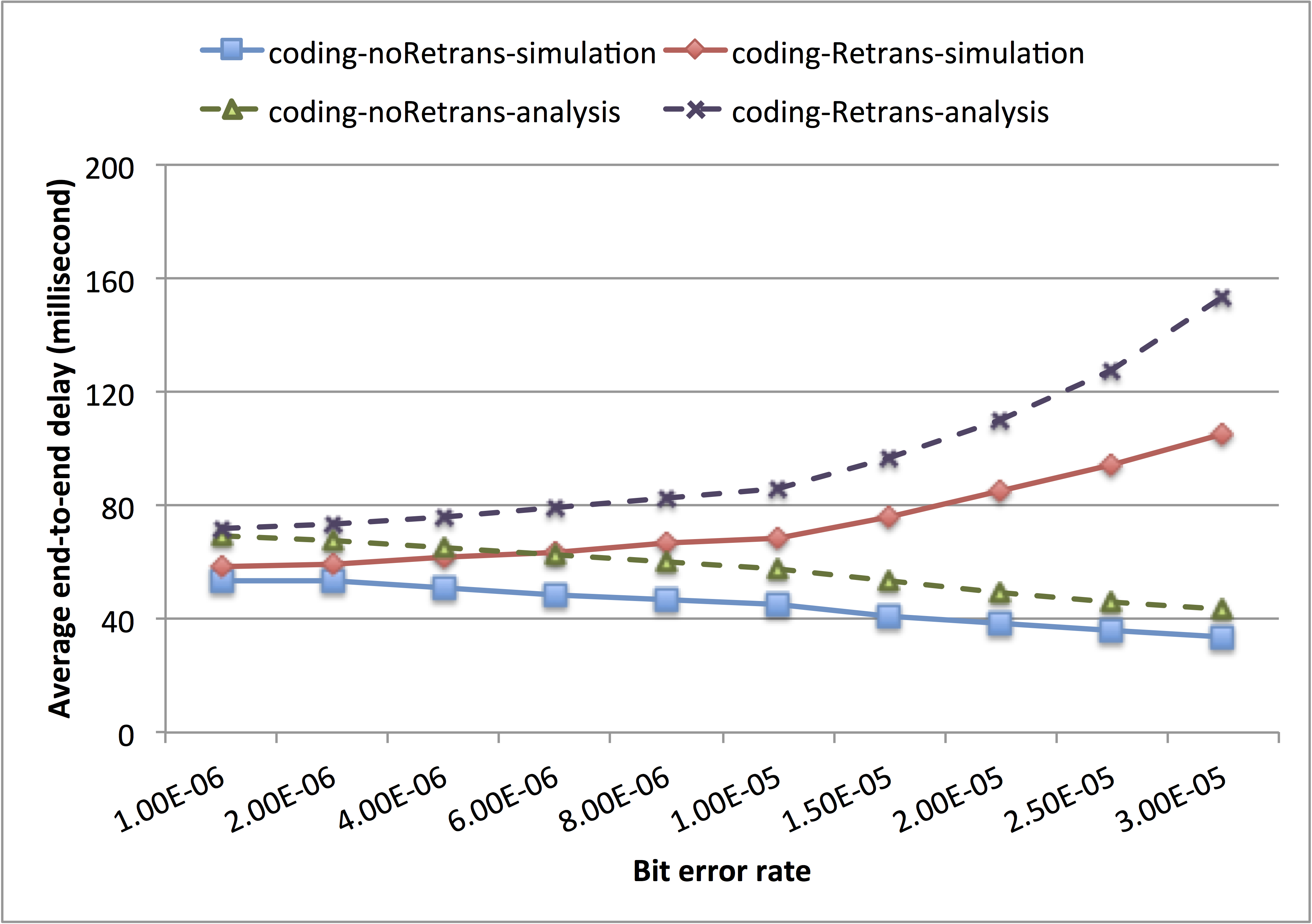}
  \caption{Coding scheme.}
  \label{fig:delay-BER-coding}
\end{subfigure}
\caption{The average end-to-end delay comparison for different bit error rates in a chain topology with 5 nodes and $\gamma = 20$.}
\label{fig:delay-BER}
\end{figure*}

To study the validity of our model under different link qualities and packet loss probabilities, we change the bit error rate, and provide simulation and analytical results for coding and non-coding schemes for the same topology depicted in \figurename~\ref{fig:chain-5}, both with and without retransmission. In these experiments, the packet generation rate at both sources (i.e., $N_1$ and $N_5$) is set to 20 packets/second. In general, as shown in \figurename~\ref{fig:throughput-BER} and \figurename~\ref{fig:delay-BER}, at lower bit error rates, the network performance with retransmission is very close to the case that retransmission is disabled. This is because at higher link qualities most of the packets are delivered to the next hop with one transmission without any need to retransmission.

As shown in \figurename~\ref{fig:throughput-BER}, the throughput calculated based on the proposed model perfectly matches the simulation results for different bit error rates. In addition, when retransmission is disabled, the throughput drops with increasing the bit error rate. On the other hand by enabling retransmission, the throughput remains almost constant especially for the non-coding scheme. The reason is that retransmission provides each packet with up to $\beta$ chances to be delivered to the next-hop, which is usually sufficient for most packets in these scenarios even at higher bit error rates. One may notice that the coding scheme does not seem as resilient as the non-coding scheme when retransmission is enabled; the reason is that, to decode each coded packet, two packets should be delivered successfully rather than one, which reduces the chance of successful delivery of coded packets even when retransmission is enabled.

Regarding the average end-to-end delay, as shown in \figurename~\ref{fig:delay-BER}, when the retransmission is disabled, the delay decreases for higher bit error rates. The reason is that more packets are dropped, and dropped packets do not contribute to the delay calculations. In addition, by increasing the packet loss rate, the number of packets waiting in the transmission queue of nodes decreases, which again causes a shorter end-to-end delay for delivered packets. On the other hand, the delay increases with the bit error rate when the retransmission mechanism is utilized, as packets require more retransmissions to get to the next-hop; this adds to both service time and waiting time. 

Comparing the coding and non-coding schemes, the effect of the bit error rate is less on coding scheme than on non-coding scheme because in the coding scheme more packets can be forwarded in each transmission. Moreover, as shown in \figurename~\ref{fig:delay-BER}, the average end-to-end delay calculated based on our analytical model provides an upper bound for the simulation results in all scenarios. 

\subsection{Maximum Stable Throughput} \label{subsec:mst}

\begin{figure}[ht]
\centering
\includegraphics[scale=0.65]{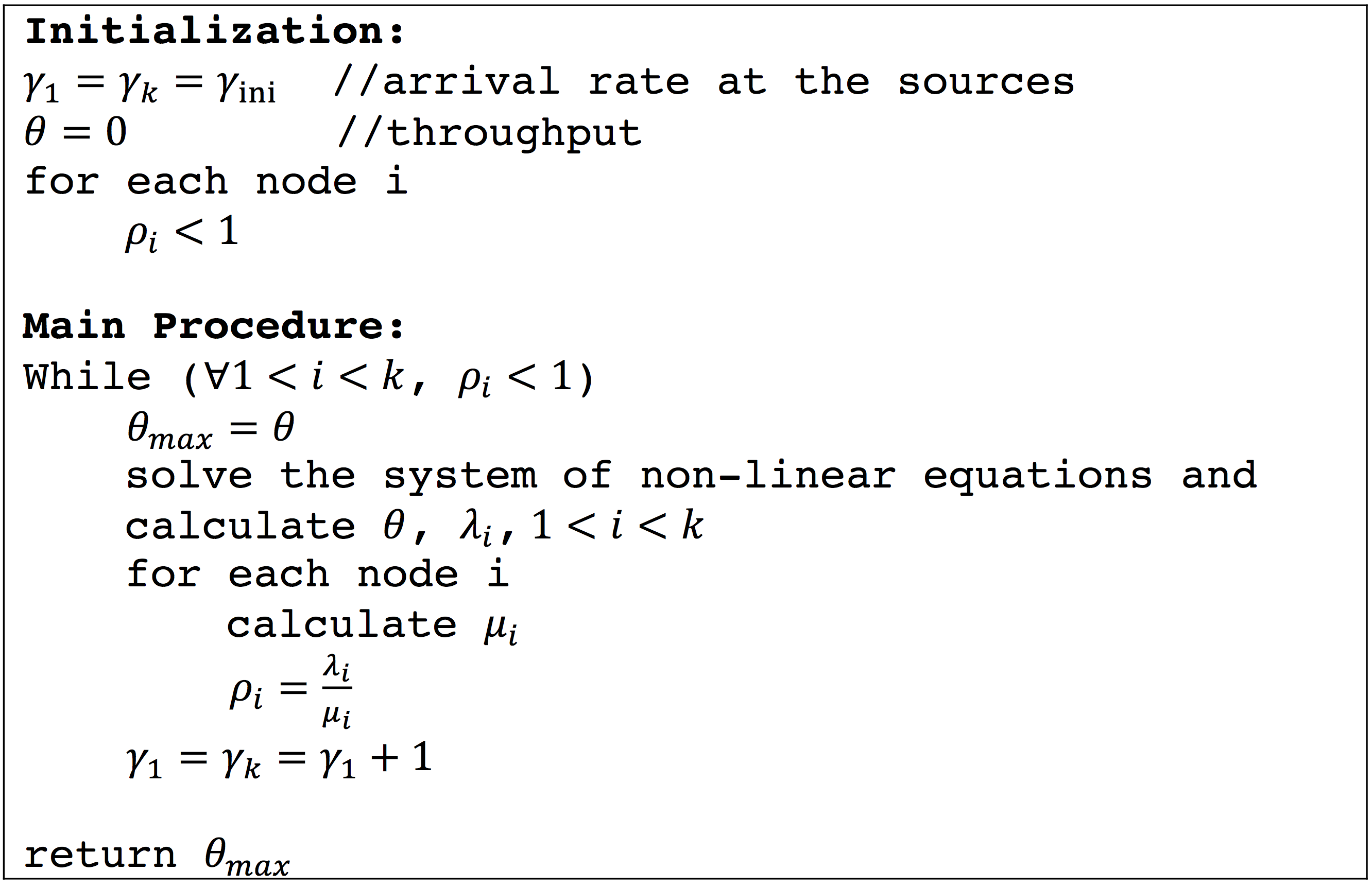}
\caption{Pseudo-code of calculating the maximum stable throughput. $\gamma_1$ and $\gamma_k$ represent the packet generation rates at the sources, initialized with a small value $\gamma_\text{ini}$. $\theta$ denotes the calculated throughput for the given generation rate.}
\label{fig:MST-pseudo}
\end{figure}

\begin{figure*}
\centering
\begin{subfigure}{.5\textwidth}
  \centering
  \includegraphics[width=0.95\linewidth]{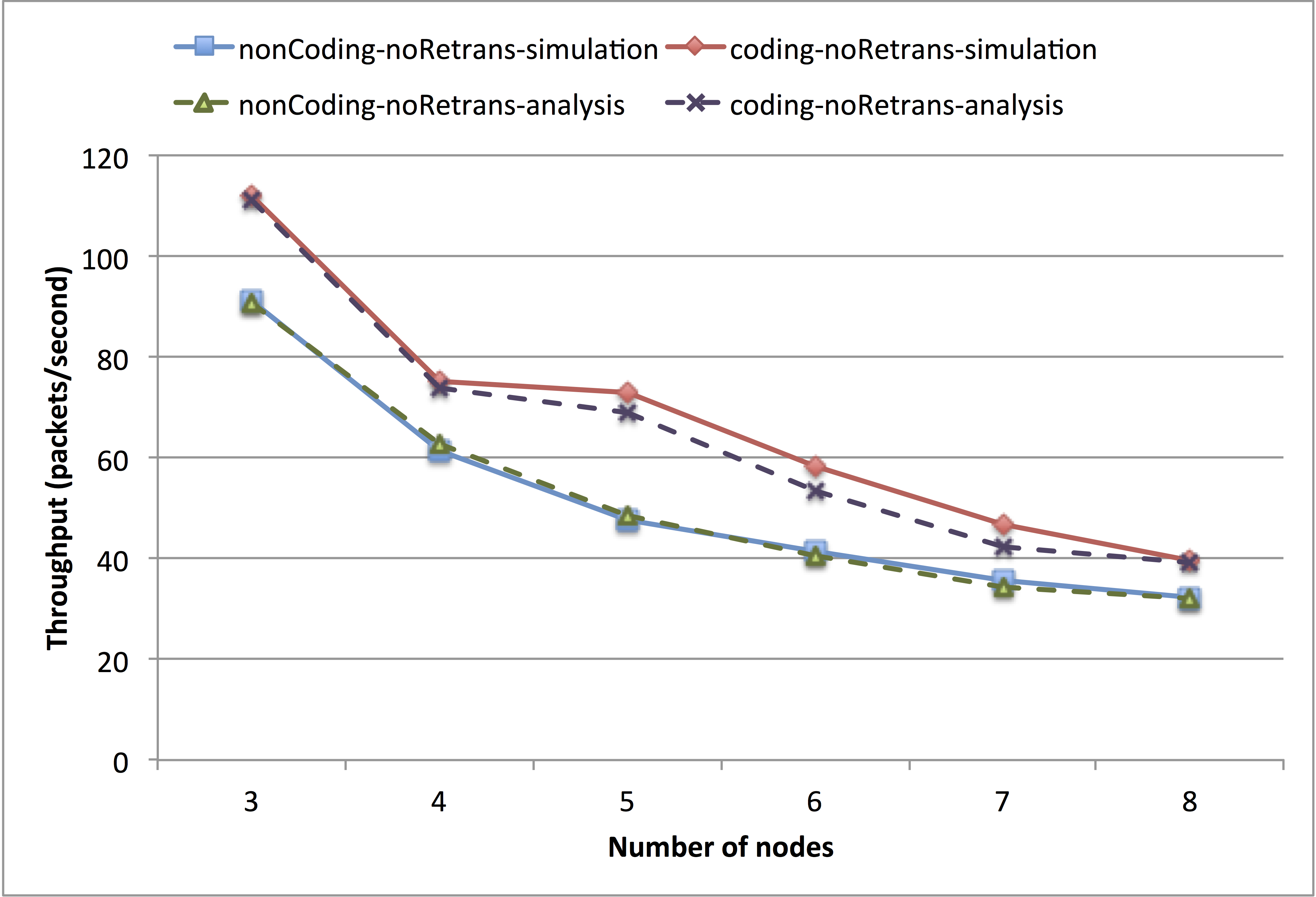}
  \caption{Without retransmission.}
  \label{fig:MST-noRetrans}
\end{subfigure}%
\begin{subfigure}{.5\textwidth}
  \centering
  \includegraphics[width=0.95\linewidth]{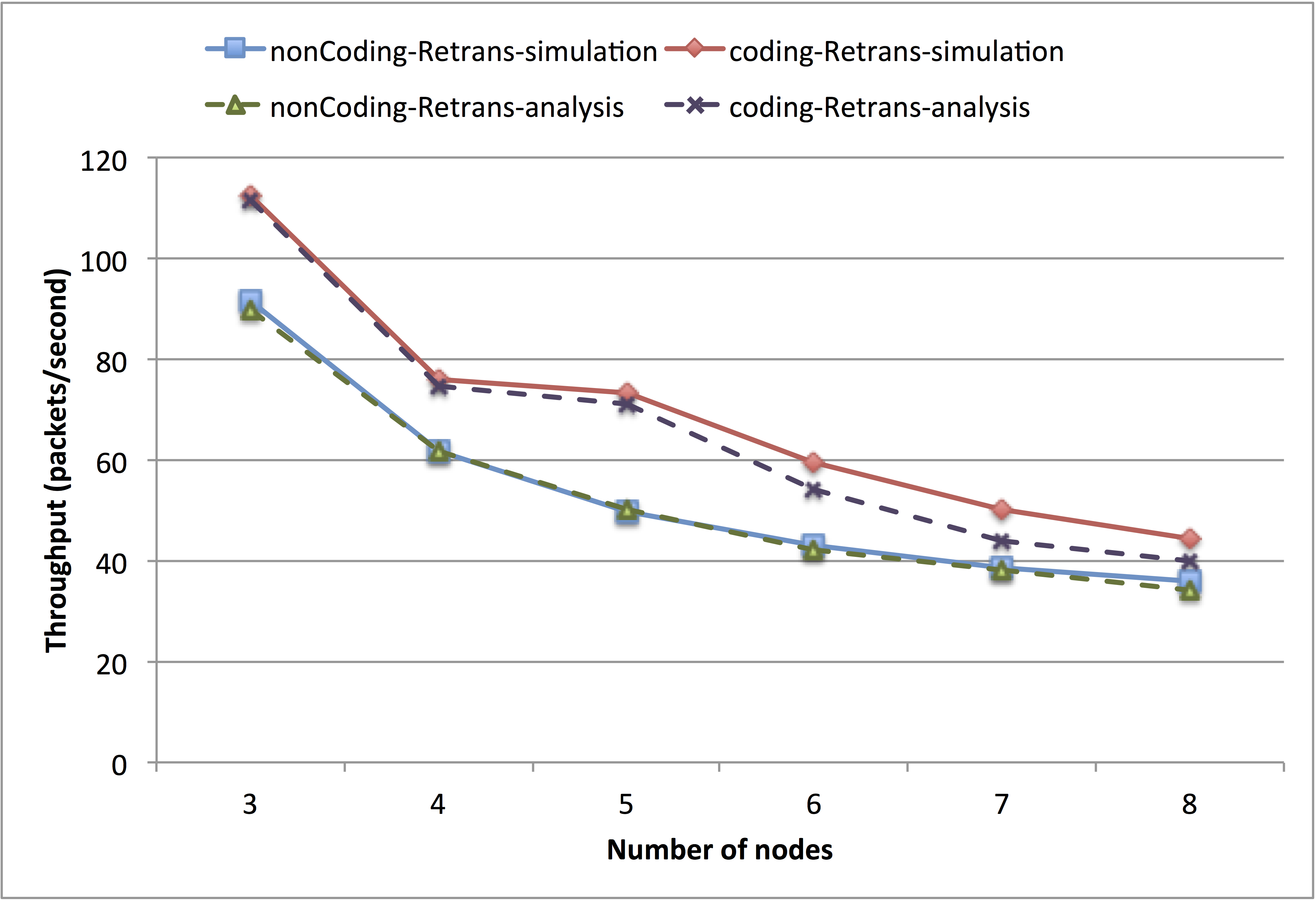}
  \caption{With retransmission.}
  \label{fig:MST-retrans}
\end{subfigure}
\caption{The maximum stable throughput comparison for different chain topology sizes, $p_e=2 \times 10^{-6}$.}
\label{fig:MST}
\end{figure*}

In this subsection, we compare the maximum stable throughput of the coding and non-coding schemes using both analytical and simulation results. The maximum stable throughput, as the name suggests, presents the maximum throughput of the network while the nodes' queues are still in a stable state (i.e., the arrival rate is less than the service rate). In these experiments, the bit error rate is set to $2 \times 10^{-6}$, and the results are provided for the chain topology depicted in \figurename~\ref{fig:chain} with variant number of nodes. 

To find the maximum stable throughput in simulations for each network size, we increase the packet generation rate at the sources as the throughput increases, and the queues are in stable state. In our analytical model, we follow the same idea since the maximum stable throughput is an increasing function of the packet generation rate. We gradually increase the packet generation rate at the sources. For each given generation rate, the system of non-linear equations provided in Section~\ref{sec:problemFormulation} is solved, providing us with the arrival rates at all nodes as well as other required parameters. Then by calculating the service rates, we can verify whether all nodes are still in a stable state. As soon as the condition of stability is not valid in at least one node, the packet generation rate is not acceptable at the sources anymore, and we calculate the throughput for the greatest acceptable generation rate as the maximum stable throughput. \figurename~\ref{fig:MST-pseudo} presents the pseudo-code for finding the maximum stable throughput in our analytical model.

As shown in \figurename~\ref{fig:MST}, our analytical model provides a good estimate of the maximum stable throughput of the network for both coding and non-coding schemes in chain topologies with different sizes. \figurename~\ref{fig:MST}-a presents the results when the retransmission mechanism is disabled, while in \figurename~\ref{fig:MST}-b, nodes are allowed to transmit each packet at most $\beta$ times. In both cases, by increasing the number of nodes in the topology, the maximum stable throughput decreases, especially in smaller topologies. In our chain topology, the number of nodes in the carrier sensing range of a transmitter is between 2 and 4, depending on the transmitter's location. As the chain length increases, a larger fraction of the nodes will have 4 nodes in their carrier sensing range, which leads to more waiting due to CSMA random access. This causes a longer back-off time and consequently a longer service time, reducing the maximum stable throughput.

Furthermore, as also stated in \cite{AnaNCQ-Amerimehr-MC2014}, when the number of intermediate nodes increases, network coding's advantage over traditional routing fades out, and the maximum stable throughput of the coding scheme approaches that of the non-coding scheme. One reason is that most coding opportunities are provided by the middle node, where the arrival rate of packets from both directions are similar and balanced. As the chain topology grows (i.e., the number of hops increases), less packets from both directions can be received by the middle nodes, which reduces the coding opportunities. In addition, in longer chains, the ratio of unbalanced flows increases in other nodes, which further causes less coding opportunities.

\section{Conclusion and Future Work} \label{sec:conclusion}
In this article, we utilized queuing theory to study the throughput and end-to-end delay of both traditional forwarding (i.e., non-coding scheme) and inter-flow network coding in multi-hop wireless mesh networks, where two unicast sessions in opposite directions traverse the network. We proposed an analytical framework considering the specifications of the IEEE 802.11 DCF, such as the binary exponential back-off time with clock freezing and virtual carrier sensing, to formulate the links quality, waiting time of the packets and retransmissions. Our analytical model assumes M/M/1 queues, which are in a stable state, while coded and native packets arrive at separated queues and coded packets have a non-preemptive higher priority over native packets. Furthermore, in our model as opposed to previous studies, the transmission of native packets is not artificially delayed for generating more coded packets; this makes it significantly more challenging to estimate coding opportunities at nodes, as described in Section~\ref{subsubsec:coding module}. 

We verified the accuracy of the proposed analytical model by computer simulation in NS-2, and the consistency of the results corroborates the validity of the model. Also, the results show that at any given packet generation rate, both throughput and end-to-end delay are higher when retransmission is enabled. However, when the bit error rate increases, the trend is totally different with and without retransmission. By enabling retransmission, throughput stays almost constant across different bit error rates while the end-to-end delay increases significantly. On the other hand, when retransmission is disabled, both throughput and end-to-end delay are decreasing functions of the bit error rate. In addition, while network coding in theory promises a greater capacity for wireless networks, the results for the maximum stable throughput show that when PHY/MAC layer constraints are taken into account, this promise can be fulfilled better for smaller topologies. In fact, when the number of intermediate nodes increases, the maximum stable throughput of network coding becomes comparable to traditional forwarding. However, wireless mesh networks are meant as an extended access technology, and it is unlikely to have very long paths; thus network coding can still offer a competitive edge.

Although our analytical model was formulated in a chain topology, it is applicable to any topology as long as the two opposite flows follow the same path. A future extension of our work could be to develop an analytical framework for a general topology, where more than two flows are traveling and possibly mixing together. In addition, we plan to incorporate cooperative forwarding to our model, where the neighbors of the next-hop can forward the packet if the next-hop itself does not receive it. 

Physical-layer network coding (PNC)~\cite{PNC-Zhang-MobiCom2006, PNC-Popovski-ICC2006, PNC-Liew-PhyComm2013} is a more recent type of network coding, in which nodes simultaneously transmit packets to a relay node that exploits mixed wireless signals to extract a coded packet. In recent years, a number of analytical studies have investigated the throughput capacity of PNC in multi-hop networks~\cite{PNC2hop-Lin-TWC2013, TwoHop-Lin-TWC2016, physicalMultiHop-Lin-TWC2016}, and we believe the model proposed here to study the throughput and delay of chain topologies can be extended into PNC, where some two-hop nodes can transmit simultaneously to a relay node without causing collision but these concurrent transmissions increase the carrier sensing range of the network~\cite{physicalMultiHop-Lin-TWC2016}.

\ifCLASSOPTIONcaptionsoff
  \newpage
\fi

% trigger a \newpage just before the given reference
% number - used to balance the columns on the last page
% adjust value as needed - may need to be readjusted if
% the document is modified later
%\IEEEtriggeratref{8}
% The "triggered" command can be changed if desired:
%\IEEEtriggercmd{\enlargethispage{-5in}}

% references section

% can use a bibliography generated by BibTeX as a .bbl file
% BibTeX documentation can be easily obtained at:
% http://mirror.ctan.org/biblio/bibtex/contrib/doc/
% The IEEEtran BibTeX style support page is at:
% http://www.michaelshell.org/tex/ieeetran/bibtex/
%\bibliographystyle{IEEEtran}
% argument is your BibTeX string definitions and bibliography database(s)
%\bibliography{IEEEabrv,../bib/paper}
%
% <OR> manually copy in the resultant .bbl file
% set second argument of \begin to the number of references
% (used to reserve space for the reference number labels box)
%\begin{thebibliography}{1}

\bibliographystyle{IEEEtran} % this is one type of author-year style
%\balance
\bibliography{citation} % this prints the bibliography section based on the \cite commands

%\bibitem{IEEEhowto:kopka}
%H.~Kopka and P.~W. Daly, \emph{A Guide to \LaTeX}, 3rd~ed.\hskip 1em plus
 % 0.5em minus 0.4em\relax Harlow, England: Addison-Wesley, 1999.

%\end{thebibliography}

% biography section
% 
% If you have an EPS/PDF photo (graphicx package needed) extra braces are
% needed around the contents of the optional argument to biography to prevent
% the LaTeX parser from getting confused when it sees the complicated
% \includegraphics command within an optional argument. (You could create
% your own custom macro containing the \includegraphics command to make things
% simpler here.)

\begin{IEEEbiography}[{\includegraphics[width=1in,height=1.25 in,clip,keepaspectratio]{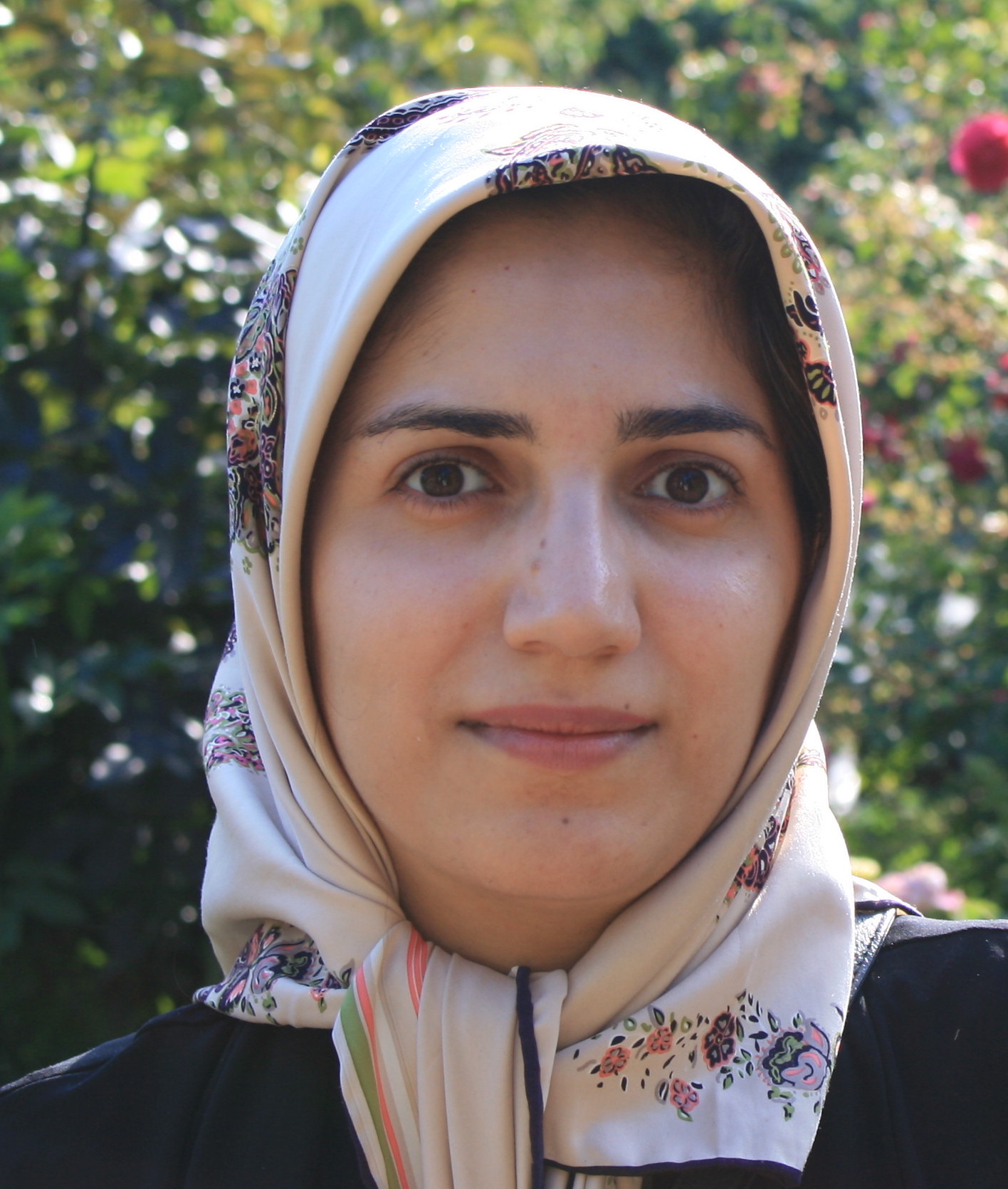}}]{Somayeh Kafaie} received her B.Sc. in Computer Engineering (Software) from Amirkabir University of Technology, Iran and her M.Sc. in Computer Engineering (Software) from Iran University of Science and Technology, Iran, in 2007 and 2011, respectively. She is currently a Ph.D. candidate in the faculty of Engineering and Applied Science, Memorial University of Newfoundland. Her research interests include wireless mesh networks, network coding, opportunistic routing, complex networks and graph theory. 
\end{IEEEbiography}

\begin{IEEEbiography}[{\includegraphics[width=1in,height=1.25 in,clip,keepaspectratio]{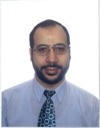}}]{Mohamed Hossam Ahmed} obtained his Ph.D. degree in Electrical Engineering in 2001 from Carleton University, Ottawa, where he worked from 2001 to 2003 as a senior research associate. In 2003, he joined the Faculty of Engineering and Applied Science, Memorial University where he works currently as a Full Professor. Dr. Ahmed published more than 135 papers in international journals and conferences. He serves as an Editor for IEEE Communication Surveys and Tutorials and as an Associate Editor for Wiley International Journal of Communication Systems and Wiley Communication and Mobile Computing (WCMC). He served as a Guest Editor of a special issue on Fairness of Radio Resource Allocation, EURASIP JWCN in 2009, and as a Guest Editor of a special issue on Radio Resource Management in Wireless Internet, Wiley Wireless and Mobile Computing Journal, 2003. Dr. Ahmed is a Senior Member of the IEEE. He served as a cochair of the Signal Processing Track in ISSPIT’14 and served as a cochair of the Transmission Technologies Track in VTC’10-Fall, and the multimedia and signal processing symposium in CCECE’09. Dr. Ahmed won the Ontario Graduate Scholarship for Science and Technology in 1997, the Ontario Graduate Scholarship in 1998, 1999, and 2000, and the Communication and Information Technology Ontario (CITO) graduate award in 2000. His research interests include radio resource management in wireless networks, multi-hop relaying, cooperative communication, vehicular ad-hoc networks, cognitive radio networks, and wireless sensor networks. Dr. Ahmed’s research is sponsored by NSERC, CFI, QNRF, Bell/Aliant and other governmental and industrial agencies. Dr. Ahmed is a registered Professional Engineer (P.Eng.) in the province of Newfoundland, Canada. 
\end{IEEEbiography}

\begin{IEEEbiography}[{\includegraphics[width=1in,height=1.25 in,clip,keepaspectratio]{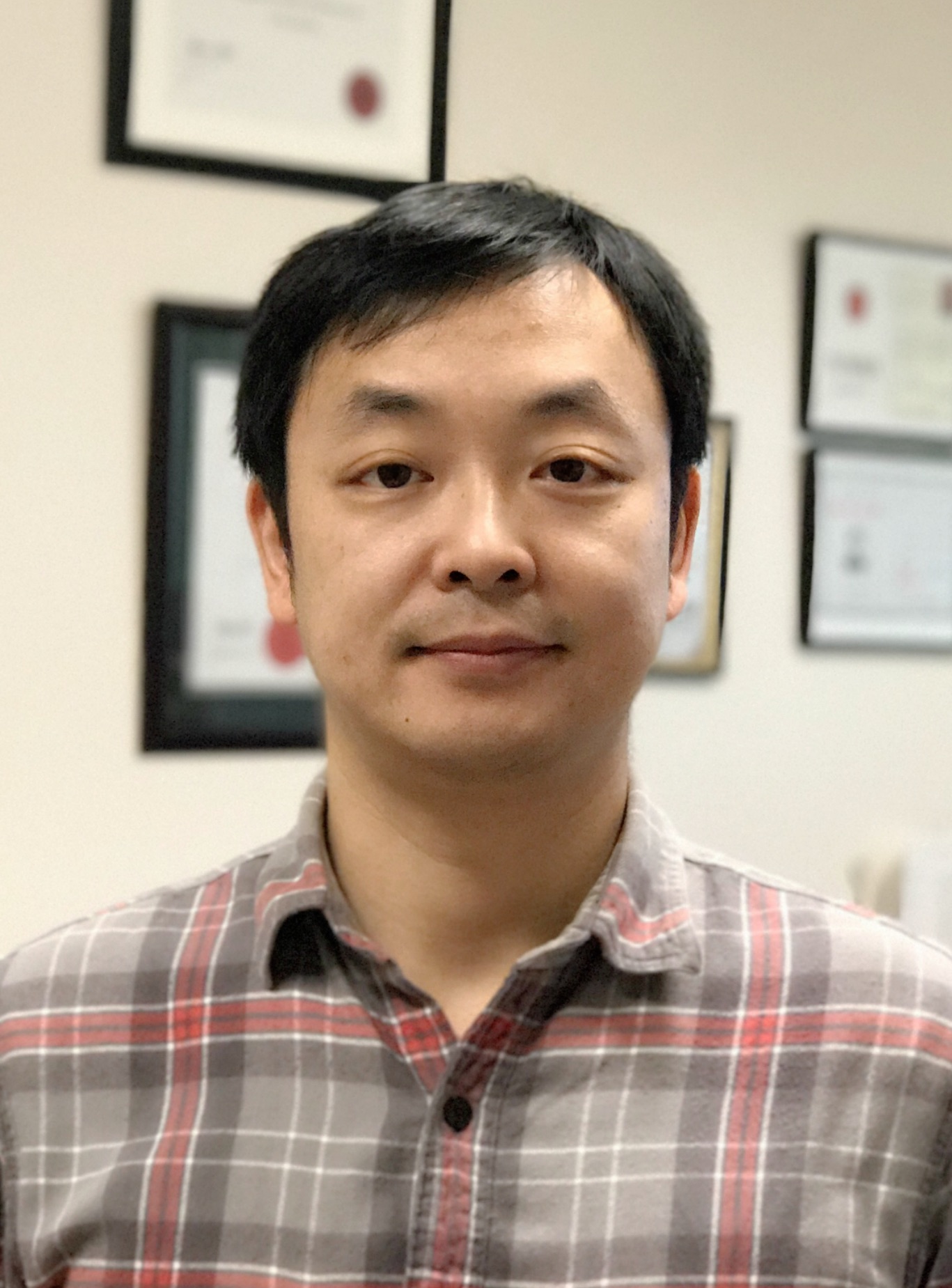}}]{Yuanzhu Chen} is an Associate Professor in the Department of Computer Science at Memorial University of Newfoundland, St. John's, Newfoundland. He was Deputy Head for Undergraduate Studies in 2012-2015, and Deputy Head for Graduate Studies in 2016 to present date.  He received his Ph.D. from Simon Fraser University in 2004 and B.Sc. from Peking University in 1999. Between 2004 and 2005, he was a post-doctoral researcher at Simon Fraser University. His research interests include computer networking, mobile computing, graph theory, complex networks, Web information retrieval, and evolutionary computation. 
\end{IEEEbiography}

\begin{IEEEbiography}[{\includegraphics[width=1in,height=1.25 in,clip,keepaspectratio]{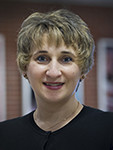}}]{Octavia A. Dobre} (M’05–SM’07) received the Engineering Diploma and Ph.D. degrees from Politehnica University of Bucharest (formerly Polytechnic Institute of Bucharest), Romania, in 1991 and 2000, respectively. She was the recipient of a Royal Society Scholarship at Westminster University (2000), and held a Fulbright Fellowship at Stevens Institute of Technology, USA (2001). Between 2002 and 2005 she was with Politehnica University of Bucharest and New Jersey Institute of Technology, USA. In 2005 she joined Memorial University, Canada, where she is currently is a Full Professor and Research Chair. Dr. Dobre was a Visiting Professor at the Université de Bretagne Occidentale, France, and Massachusetts Institute of Technology (MIT), USA (2013).  
Her research interests include 5G technologies, blind signal identification and parameter estimation techniques, cognitive radio systems, and transceiver optimization algorithms for wireless communications, as well as optical and underwater communications. She has co-authored over 190 journal and conference papers and gave more than 40 invited talks to industry and academia.
%Her research has been supported by the Natural Sciences and Engineering Research Council of Canada, Mathematics of Information Technology and Complex Systems, Canada Foundation for Innovation, Research and Development Corporation, Atlantic Canada Opportunities Agency, Defence and Research Development Canada, Communications Research Centre Canada, Altera Corporation, Allen Vanguard, DTA Systems, EION Wireless, ThinkRF, and Agilent Technologies. 

Dr. Dobre serves as the Editor-in-Chief of the IEEE Communications Letters, as well as an Editor for the IEEE Communications Surveys and Tutorials and IEEE Systems. She was an Editor and a Senior Editor for the IEEE Communications Letters, Editor for the IEEE Transactions on Wireless Communications and Guest Editor for other prestigious journals. She served as General Chair of CWIT, and Technical Co-Chair of symposia at numerous conferences, such as IEEE GLOBECOM and ICC. She is the Chair of the IEEE ComSoc Signal Processing for Communications and Electronics Technical Committee, the Chair of the IEEE ComSoc WICE standing committee, as well as a member-at-large of the Administrative Committee of the IEEE Instrumentation and Measurement Society. 
\end{IEEEbiography}

%\begin{IEEEbiography}[{\includegraphics[width=1in,height=1.25in,clip,keepaspectratio]{mshell}}]{Michael Shell}
% or if you just want to reserve a space for a photo:

%\begin{IEEEbiography}{Michael Shell}
%Biography text here.
%\end{IEEEbiography}

% if you will not have a photo at all:
%\begin{IEEEbiographynophoto}{John Doe}
%Biography text here.
%\end{IEEEbiographynophoto}

% insert where needed to balance the two columns on the last page with
% biographies
%\newpage

%\begin{IEEEbiographynophoto}{Jane Doe}
%Biography text here.
%\end{IEEEbiographynophoto}

% You can push biographies down or up by placing
% a \vfill before or after them. The appropriate
% use of \vfill depends on what kind of text is
% on the last page and whether or not the columns
% are being equalized.

%\vfill

% Can be used to pull up biographies so that the bottom of the last one
% is flush with the other column.
%\enlargethispage{-5in}

% that's all folks
\end{document}